\documentclass[10pt]{article}

\usepackage{amsmath}

\usepackage{amssymb}

\usepackage{graphicx}

\usepackage{cite}

\usepackage{color}
\definecolor{MyMagenta}{cmyk}{0.,1.,0.,0.}
\definecolor{MyBlue}{rgb}{0.,0.,1.}
\definecolor{MyRed}{rgb}{1.,0.,0.}

\usepackage[labelfont=bf,labelsep=period,justification=raggedright]{caption}

\makeatletter
\renewcommand{\@biblabel}[1]{\quad#1.}
\makeatother

\newcommand{\DemianRev}[1]{#1}%{\textcolor{MyBlue}{\textbf{#1}}}
\newcommand{\OlavRev}[1]{#1}%{\textcolor{MyBlue}{\textbf{#1}}}
\newcommand{\JordiRev}[1]{#1}%{\textcolor{MyBlue}{\textbf{#1}}}

\date{}

\begin{document}

% Title must be 150 characters or less
\begin{flushleft}
    {\Large \textbf{Model-Free Reconstruction of \DemianRev{Excitatory Neuronal} Connectivity from Calcium Imaging Signals} } \\
    Olav Stetter$^{1,2,3}$, Demian Battaglia$^{1, 3,\ast}$, Jordi Soriano$^{4}$, Theo Geisel$^{1,2,3}$ \\
    \textbf{1} Max Planck Institute for Dynamics and Self-Organization, G\"ottingen, Germany \\
    \textbf{2} Georg August University, Physics department, G\"ottingen, Germany \\
    \textbf{3} Bernstein Center for Computational Neuroscience, G\"ottingen, Germany \\
    \textbf{4} Departament d'ECM , Facultat de F\'{\i}sica, Universitat de Barcelona, Barcelona, Spain \\
    $\ast$ E-mail: demian$@$nld.ds.mpg.de
\end{flushleft}

% \emph{(draft version 17, compiled on \today)}

% Please keep the abstract between 250 and 300 words
\section*{Abstract}

A systematic assessment of global neural network connectivity through direct electrophysiological assays has remained technically infeasible, even in dissociated neuronal cultures. We introduce an improved algorithmic approach based on Transfer Entropy to reconstruct approximations to structural connectivity from network activity monitored through calcium imaging. \DemianRev{We focus in this study, as a first step, on the inference of excitatory synaptic links.} Based on information theory, our method requires no prior assumptions on the statistics of neuronal firing and neuronal connections. The performance of our algorithm is benchmarked on surrogate time series of calcium fluorescence generated by the simulated dynamics of a network with known ground truth topology. We find that the \DemianRev{functional} network topology revealed by Transfer Entropy depends qualitatively on the time-dependent dynamic state of the network (bursting or non-bursting). Thus by conditioning with respect to the global mean activity, we improve the performance of our method. This allows us to focus the analysis to specific dynamical regimes of the network in which the inferred functional connectivity is shaped by the underlying \DemianRev{excitatory connections}, rather than by collective synchrony. Our method can discriminate between actual causal influences between neurons and spurious non-causal correlations due to light scattering artifacts, which inherently affect the quality of fluorescence imaging. Compared to other reconstruction strategies such as cross-correlation or Granger Causality methods, our method based on improved Transfer Entropy is remarkably more accurate. In particular, it provides a good reconstruction of the \DemianRev{excitatory} network clustering coefficient, allowing for discrimination between weakly and strongly clustered topologies. Finally, we present the applicability of our method to real recordings of \emph{in vitro} cortical cultures. We demonstrate that \DemianRev{excitatory connections in} these networks are characterized by an elevated level of clustering compared to a random graph (although not extreme) and \DemianRev{can be} markedly non-local.
%%%%% 296 WORDS

% Please keep the Author Summary between 150 and 200 words
% Use first person. PLoS ONE authors please skip this step.
% Author Summary not valid for PLoS ONE submissions.
\section*{Author Summary} 
Unraveling general organization principles of connectivity in neural circuits is a crucial step towards understanding brain function. However, even the simpler task of assessing the global \DemianRev{excitatory} connectivity of a culture \emph{in vitro}, where neurons form self-organized networks in absence of external stimuli, remains challenging. Neuronal cultures undergo spontaneous switching between episodes of synchronous bursting and quieter inter-burst periods. We introduce here a novel algorithm which aims at inferring the connectivity of neuronal cultures from calcium fluorescence recordings of their network dynamics. To achieve this goal, we develop a suitable generalization of Transfer Entropy, an information-theoretic measure of causal influences between time series. Unlike previous algorithmic approaches to reconstruction, Transfer Entropy is data-driven and does not rely on specific assumptions about neuronal firing statistics or network topology. We generate simulated calcium signals from networks with controlled ground-truth topology \DemianRev{and purely excitatory interactions} and show that, by restricting the analysis to inter-bursts periods, Transfer Entropy robustly achieves a good reconstruction performance for disparate network connectivities. We apply, finally, our method to real data and find evidence of non-random features in cultured networks, such as the existence of high-connectivity hub \DemianRev{excitatory} neurons and of an elevated (but not extreme) level of clustering.

%%%%% 200 WORDS

\section*{Introduction}

The identification of the topological features of neuronal circuits is an essential step towards the understanding of neuronal computation and function. Despite considerable progress in \JordiRev{neuroanatomy, electrophysiology and imaging ~\cite{Sporns:2005p752, Yoshimura:2005hr, Arenkiel:2009cf,Scanziani:2009na,Gong:2009fo, Lichtman:2011kh, Perin:2011kp, Wedeen30032012}}, the detailed mapping of neuronal circuits is already a \JordiRev{difficult task} for a small population of neurons, and becomes unpractical when accessing large neuronal ensembles. Even in the case of cultures of dissociated neurons, in which neuronal connections develop \emph{de novo} during the formation and maturation of the network, very few details are known about the statistical features of this connectivity, which might reflect signatures of self-organized critical activity \cite{Levina:2007ev, Millman:2010jo, Tetzlaff:2010he}. 

Neuronal cultures have emerged in the last years as simple, yet versatile model systems \cite{Eckmann:2007ft,Feinerman:2008vc} in the quest for uncovering neuronal connectivity \cite{Soriano:2008p747,Erickson20081} and dynamics \cite{Opitz:2002fc,Marom:2002va,Wagenaar:2006fo,Cohen:2010iq}. The fact that relatively simple cultures already exhibit a rich repertoire of spontaneous activity \cite{Wagenaar:2006fo,Cohen:2008cv} make them particularly appealing for studying the interplay between activity and connectivity. 

The activity of hundreds to thousands of cells in \emph{in~vitro} cultured neuronal networks can be simultaneously monitored using calcium fluorescence imaging techniques \cite{Stosiek:2003ab,Soriano:2008p747,Grienberger2012}. \DemianRev{Calcium imaging can be applied} \JordiRev{both \emph{in~vitro} and \emph{in~vivo}} \DemianRev{and can be combined} \JordiRev{with stimulation techniques like optogenetics \cite{Yizhar:2011jv}}.
A major drawback of this technique, however, is that the typical frame rate during acquisition is slower than the cell's firing dynamics \DemianRev{by an order of magnitude. Furthermore the poor signal-to-noise ratio is such to make hard the detection of elementary firing events.} 

%\DemianRev{On the contrary, } other techniques such as multielectrode arrays \cite{Eckmann:2007ft,Marom:2002va,Beggs:2003uv,Wagenaar:2006fo,Erickson20081} \DemianRev{allow to discriminate single spikes}, but are restricted to \DemianRev{considerably} smaller populations of neurons. 

%\JordiRev{Additionally, the source of the spiking events can be often difficult to resolve since neurons are, in general, not located at the recording sites. Calcium imaging, despite its limitations, is a fast evolving technique that can monitor a large population of neurons, both \emph{in~vitro} and \emph{in~vivo}. In combination with stimulation techniques like optogenetics \cite{Yizhar:2011jv}, calcium imaging is nowadays revolutionizing our comprehension of neuronal activity and function \cit.} 

Neuronal cultures are unique platforms to investigate and quantify the accuracy of network reconstruction from activity data, extending analysis tools initially devised for the characterization of \OlavRev{macro-scale} functional networks \cite{Achard:2006fo, Honey:2007bb} to the \OlavRev{micro-scale} of a developing local circuit. 

Here we report a new technique to reconstruct the connectivity of a neuronal network from calcium imaging data, \DemianRev{which is} based on information theory. We use an extension of Transfer Entropy (TE)~\cite{Schreiber:2000p79,Kaiser:2002p2346,Wibral:2011gz} to extract a \DemianRev{\textit{directed functional connectivity}} network in which the presence of a directed edge between two nodes reflects a direct causal influence by the source to the target node~\cite{Aertsen:1989tg, Friston:1994p559, Bressler:2011bv}. Note that ``causal influence'' is defined operationally as ``improved predictability''~\cite{Wiener:ws,Granger:1969p4641} reflecting the fact that knowledge of the activity of one node (\DemianRev{putatively pre-synaptic}) is helpful in predicting the future behavior of another node (\DemianRev{putatively post-synaptic)}. 
%\DemianRev{Although TE can be considered a measure of effective connectivity in strict sense \cite{Battaglia:2012eg}, we prefer here using the less specific term of ``directed functional connectivity''.
 TE has previously been used to study gene regulatory networks~\cite{Wang:2006p2349}, the flow of information between auditory neurons~\cite{Gourevitch:2007p2350}, to infer \DemianRev{directed interactions} between brain areas based on EEG recordings~\cite{Vicente:2011p6560} or between different LFP frequency bands~\cite{Besserve:2010p591}, as well as for the reconstruction of the connectivity based on spike times~\cite{Garofalo:2009kb, Ito:2011wc}.

\DemianRev{Importantly,  our data-driven} approach is model-independent. \DemianRev{This contrasts with} previous approaches to network reconstruction, \DemianRev{which} were most often based on the knowledge of precise spike times~\cite{Cadotte:2008fs,Patnaik:2008us,Pajevic:2009p252,Vogelstein:2009p3029,Shandilya:2011vd,Mishchencko:2009p4301}, or explicitly assumed a specific model of neuronal activity~\cite{Vogelstein:2009p3029,Shandilya:2011vd}. \DemianRev{Our method  is thus} not constrained to linear interactions between nodes. This absence of a parametric model can be advantageous not only conceptually, where we hope to make the least amount of assumptions necessary, but also \DemianRev{practical for applications} to real data, because \DemianRev{the used statistical assumptions} might be too restrictive \DemianRev{or inappropriate}.

A problem inherent \OlavRev{to} the \DemianRev{indirect algorithmic inference} of network connectivity from real data is that the true target topology of the network is not known and that, therefore, it is difficult to assess the quality of the reconstruction. In order to characterize the behavior of our algorithm and to benchmark its potential performance, we resort \DemianRev{therefore} to synthetic calcium fluorescence time series generated by a \DemianRev{simulated cultured neural network} that exhibits realistic dynamics. Since the ``ground truth'' topology of \DemianRev{cultures \textit{in silico}} is known and arbitrarily selectable, the quality of our reconstruction can be evaluated by systematically comparing the inferred with the real network connectivities. 

\OlavRev{We use a simplified network simulation to generate surrogate imaging data,} \DemianRev{improving their realism} by the reproduction of \DemianRev{light scattering artifacts~\cite{Lichtman:2011kh}} which ordinarily affect the quality of the recording. \DemianRev{Our surrogate data reproduce as well another general feature} of the activity of neuronal cultures, namely the occurrence of temporally irregular switching between states of asynchronous activity, with a relatively weak average firing rates, and states of highly synchronous activity, commonly denoted as ``network bursts''~\cite{Eytan:2006p66,Cohen:2008cv,Eckmann:2008p77}. 

This switching dynamics poses potentially a major obstacle to reconstruction. \DemianRev{As a matter of fact,} during bursting phases, collective synchronization results in an effective network with densely connected components (reflecting communities of highly synchronized neurons). Conversely ---and usefully for applications--- the \DemianRev{directed functional} connectivity during inter-bursts phases bears a strong resemblance to the underlying \textit{structural} (i.e. synaptic) connectivity, because it reflects dominantly mono-synaptic interactions. \DemianRev{In order not to average over dynamical regimes leading to very different functional networks}, we restrict our analysis to the \DemianRev{inter-burst} dynamical regime only, \DemianRev{ignoring bursting transients}. This can be achieved very simply by \emph{conditioning} our analysis to activity intervals in which the averaged fluorescence level is below a certain threshold, as an indirect but reliable indicator of whether the network is in a comparatively ``quiet'' phase. 

\DemianRev{Appropriate conditioning ---combined with a simple correction coping with the poor time-resolution of imaging data--- allows to achieve a good topology reconstruction performance (assessed from synthetic data)}, without the need to infer exact spike times through sophisticated techniques (as is required, \DemianRev{on the contrary}, in \cite{Vogelstein:2009p3029, Mishchencko:2009p4301}). \DemianRev{As a matter of fact, TE out-performs} three other standard approaches, namely cross-correlation~(XC), Granger Causality~(GC) and Mutual Information~(MI) ---all of which have been used to study the connectivity in neural networks~\cite{Cadotte:2008fs,Garofalo:2009jm,Li:2010p5556,Singh:2010br,Ostwald:2011te,Biffi:2011kl,Ferguson:2011fj}---, \DemianRev{particularly in presence of light scattering}.

%A further problematic aspect for reconstruction based on calcium fluorescence time series is that synaptic delays are short compared to the typical sampling rate of our recordings. This makes it difficult to determine causal interactions based on the original formulation of TE (requiring the sampling of time-lagged probability distributions), because pairs of causally-related pre- and post-synaptic events are commonly confounded within a same frame of the recording. We therefore introduce here a simple but effective technical modification, taking into account ``same bin'' interactions (see {\it Materials and Methods}) and achieving good performance without the need to infer exact spike times through sophisticated deconvolution techniques (as is required, e.g., in \cite{Vogelstein:2009p3029, Mishchencko:2009p4301}).

\DemianRev{Finally}, we apply our algorithm \DemianRev{---optimized through model-based validation---} to the analysis of real calcium imaging recordings. For this purpose, we study spontaneously developing networks of dissociated cortical neurons \emph{in vitro} \DemianRev{and we address, as a first step toward a full topology reconstruction, the simpler problem of extracting their excitatory connectivity only}. 
\OlavRev{Early mature} cultures display a bursting dynamics very similar to our simulated networks, \DemianRev{with which they also share} \OlavRev{an analogous state-dependency of directed functional connectivity.}

Our generalized TE approach identifies \DemianRev{thus} network topologies \DemianRev{with characteristically non-local connections}, consistent with the existence of long axons able to span a large fraction of the culture~\cite{Kriegstein:1983uj,Feinerman:2008vc,Erickson20081}. The retrieved degree distribution is broadened and characteristically right-skewed (but not ``scale free'', in contrast with~\cite{Pajevic:2009p252}). \DemianRev{Furthermore, our analysis hints at} a level of clustering \DemianRev{in spontaneously-developing cultures} which is moderate but significantly larger than expected for random networks sharing the same degree distribution.

% Results and Discussion can be combined.
\section*{Results}

\DemianRev{The \textit{Results} section is organized as follows. After a brief presentation of the qualitative similarity
between real calcium fluorescence data from neuronal cultures and simulated data, we introduce
numerical simulations showing that networks with very different clustering levels can lead to matching
bursting dynamics. We develop then our reconstruction strategy, based on a novel generalization
of TE, and examine the different elements composing our strategy, namely \textit{``same-bin interactions''}
and \textit{conditioning} with respect to the average fluorescence level. We show that only signals
recorded during inter-burst periods convey elevated information about the underlying structural topology.
After a discussion of criteria guiding the choice of the number of links to include in the
reconstructed network, we illustrate two specific examples of reconstruction, and analyze the factors
affecting its quality. The performance achieved by generalized TE is systematically contrasted with other
standard linear and nonlinear competitor methods, all of them modified to include the same enhancements
applied to the original TE formulation. Finally, we apply our reconstruction algorithm to biological recordings
and extract the topological features of actual neuronal cultures.}

\subsection*{\DemianRev{Real and surrogate calcium fluorescence data}}

\begin{figure}
    [!ht]
    \begin{center}
        \includegraphics[width=12.35cm]{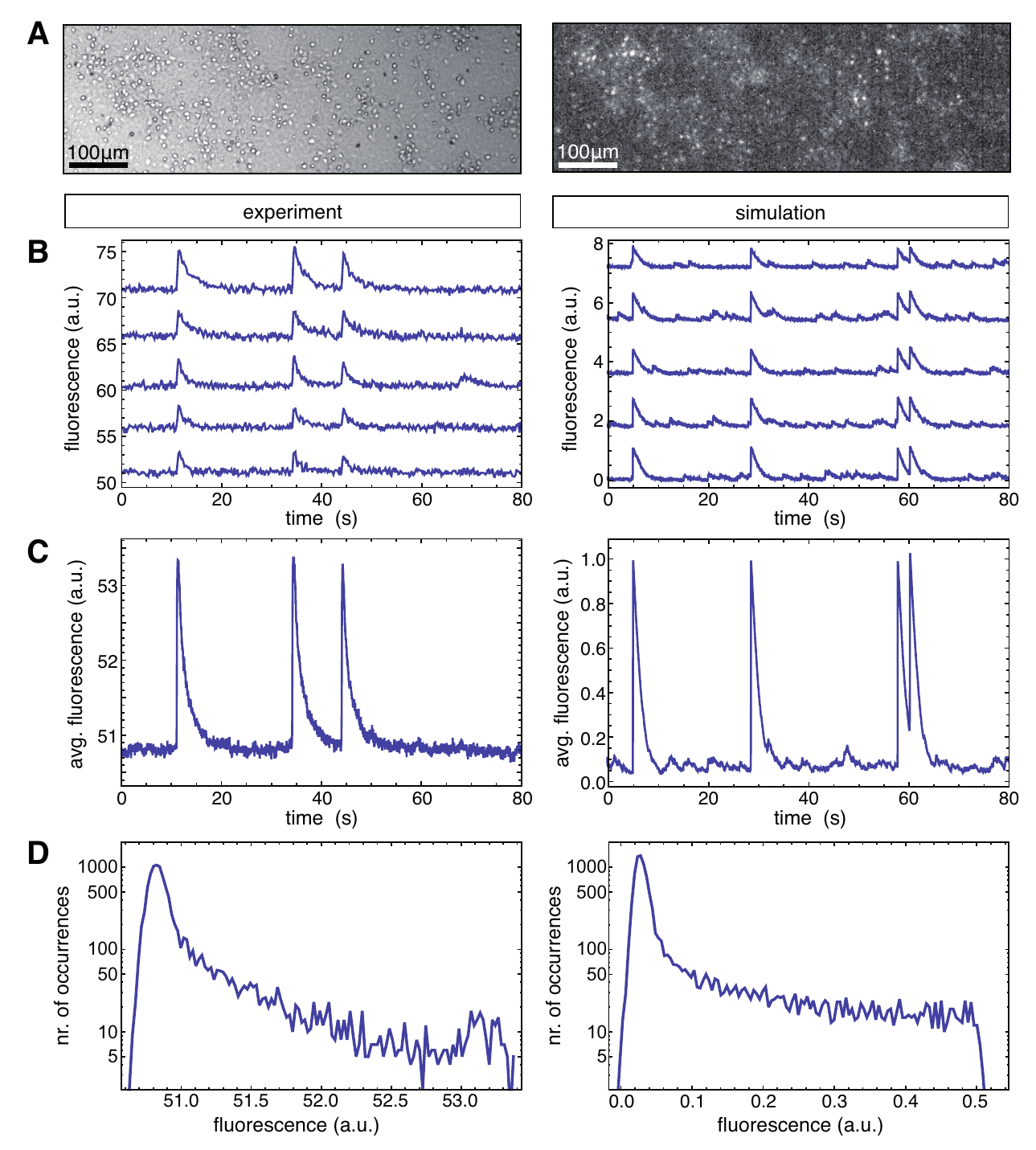}
    \end{center}
    \caption{} \label{fig:raw_data_vs_simulations}
\end{figure}

\DemianRev{In this study, we consider} recordings from \textit{in vitro} cultures of dissociated cortical neurons (see \textit{Materials and Methods}). \JordiRev{To illustrate the quality of our recordings, in Fig.~\ref{fig:raw_data_vs_simulations}A we provide a bright field image of a region of a culture together with its associated calcium fluorescence.} \DemianRev{As previously anticipated,} to simplify the network reconstruction problem, experiments are carried out with blocked inhibitory GABA-ergic transmission, so that the network activity is driven solely by excitatory connections. \DemianRev{We record activity of \textit{early mature}} cultures at day \emph{in vitro} (DIV) 9-12. \JordiRev{Such young but sufficiently mature cultures display rich bursting events,} \DemianRev{combined with sparse irregular firing activity during inter-burst periods (cfr. \textit{Discussion})}. 

\JordiRev{In Fig.~\ref{fig:raw_data_vs_simulations}B (left panel) we show actual recordings of the fluorescence traces associated to five different neurons. The corresponding population average for the same time window is shown in Fig.~\ref{fig:raw_data_vs_simulations}C (left panel). In these recordings,} a stable baseline is broken by intermittent activity peaks that correspond to \DemianRev{synchronized} network bursts \DemianRev{recruiting many neurons}. The bursts display a fast rise of fluorescence at their onset followed by a slow decay. \DemianRev{In addition, during inter-burst periods, smaller modulations above the baseline are sometimes visible, despite the poor time-resolution achievable of a frame every few tens of a ms.}

\DemianRev{In order to benchmark and optimize different reconstruction methods, we also generate surrogate calcium fluorescence data (shown in the right panels of Figs.~\ref{fig:raw_data_vs_simulations}B and \ref{fig:raw_data_vs_simulations}C), based on the activity of simulated networks whose ground truth topology is known.}
We simulate the spontaneous spiking dynamics of networks formed by $N=100$ excitatory integrate-and-fire neurons, along a duration of 60~minutes of real time, \DemianRev{matching typical lengths of actual recordings. Calcium fluorescence time series are then produced based on this spiking dynamics, resorting to a model introduced in~\cite{Vogelstein:2009p3029} and described in the \textit{Materials and Methods} section.}

Although in our experiments \JordiRev{a number of over 1000 cells is accessible}, we observed in the simulations that~$N=100$ neurons suffice to reproduce the same dynamical behavior observed for larger network sizes, while allowing still for an exhaustive exploration of the entire algorithmic parameter space. \DemianRev{Furthermore, despite their reduced density, we maintain in simulated cultures the same average probability of connection than in actual cultures, where this probability (of $p = 0.12$, see \textit{Materials and Methods}) is roughly estimated based on independent studies~\cite{Erickson20081,Wen28072009}.}

The fluorescence signal of a particular simulation run or experiment can be conveniently studied in terms of the distribution of fluorescence amplitudes. As shown in Fig.~\ref{fig:raw_data_vs_simulations}D for both simulations and experiments, the amplitude distributions display a characteristic right-skewed shape that emerge from the switching between two distinct dynamical regimes, \JordiRev{namely the presence or absence of bursts}. The distribution in the low fluorescence region assumes a Gaussian-like shape, corresponding to noise-dominated baseline activity, while the high fluorescence region displays a long tail with a cut-off at the level of calcium fluorescence of the highest network spikes. As we will show later, qualitative similarity between the shapes of the simulated and experimental fluorescence distributions will play an important guiding role for an appropriate network reconstruction.

\JordiRev{Note that we do not attempt to reproduce quantitatively the experimental distribution of fluorescence through our simplified model.} \DemianRev{We remark, however, that the right tail of our experimental and simulated distributions of fluorescence values cannot be well modeled by a power law. This contrasts with reports \cite{Beggs:2003uv,Mazzoni:2007jq} of power-laws in the distribution of burst sizes, even if the comparison might be misleading, due to the indirect relation ---although linear, at least in a certain range \cite{Takano:2012jw}--- between fluorescence and burst size.}

\subsection*{Different network topologies lead to \DemianRev{equivalent} network bursting}

Neurons grown \textit{in vitro} develop on a bi-dimensional substrate and, hence, both connectivity and clustering may be strongly sensitive to the physical distance between neurons. At the same time, due to long axonal projections~\cite{Kriegstein:1983uj,Feinerman:2008vc} \DemianRev{excitatory} synaptic connections might be formed at any distance within the whole culture and both activity \JordiRev{and signaling-dependent mechanisms} might shape non-trivially long-range connectivity~\cite{Sur:1999wo, Yu:2001ch}.

To test the reconstruction performance of our algorithm, we consider two general families of network topologies that cover a wide range of clustering coefficients. In a first one, clustering occurs between randomly positioned nodes (\textit{non-local} clustering). In a second one, the connection probability between two nodes decays with their Euclidean distance \OlavRev{according to a Gaussian distribution} and, therefore, connected nodes are also likely to be spatially close. In particular, in this latter case, the overall level of clustering is determined by how fast the connection probability decays with distance (\textit{local} clustering). Cortical slice studies revealed the existence of both local~\cite{Holmgren:2003p651,Song:2005p45} and non-local~\cite{Kalisman:2005fq, Perin:2011kp} types of clustering. 

\OlavRev{We will later benchmark reconstruction performance for both kinds of topologies and for a wide range of clustering levels, because} \DemianRev{very similar patterns of neuronal activity can be generated by very different networks}, as we now show.

\begin{figure}
    [!ht]
    \begin{center}
        \includegraphics[width=17.35cm]{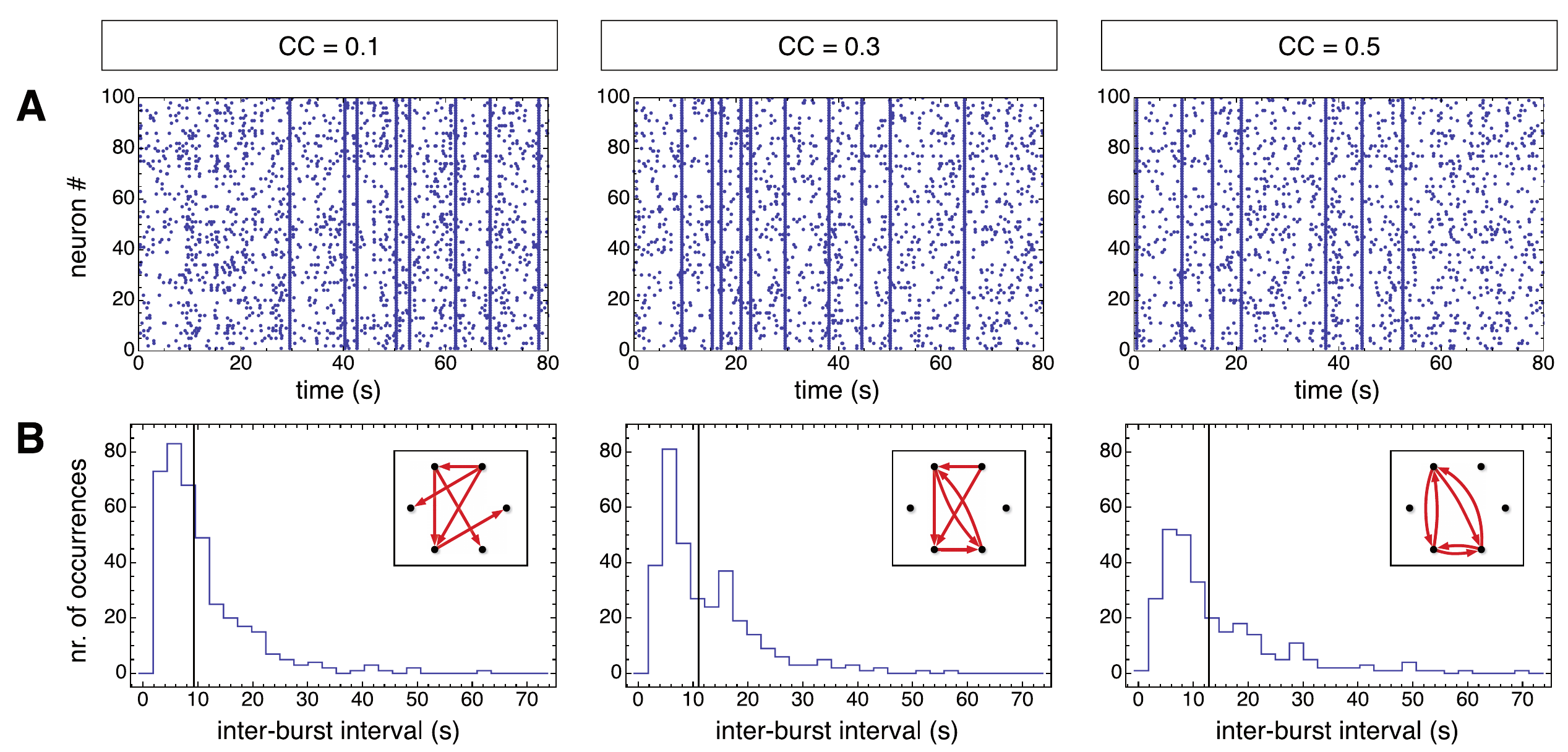}
    \end{center}
    \caption{} \label{fig:CCnets_all_look_alike}
\end{figure}

Figure~\ref{fig:CCnets_all_look_alike} illustrates the dynamic behavior of three networks (in this case from the non-local clustering ensemble). The networks are designed to have different clustering coefficients but the same total number of links (see the insets of Fig.~\ref{fig:CCnets_all_look_alike}B for an illustration). The synaptic coupling between neurons was adjusted in each network using an automated procedure to obtain bursting activities with comparable bursting rates (see Methods for details and Table~\ref{tab:synaptic_weights_used_for_simulation} for the actual values of the synaptic weight). As a net effect of this procedure, the synaptic coupling between neurons is slightly reduced for larger clustering coefficients. The simulated spiking dynamics is shown in the raster plots of Fig.~\ref{fig:CCnets_all_look_alike}A. 
These three networks display indeed very similar bursting dynamics, \OlavRev{not only in terms of the mean bursting rate, but also in terms of the entire inter-burst interval (IBIs) distribution, shown in Fig.~\ref{fig:CCnets_all_look_alike}B. In the same manner, we} constructed and simulated \textit{local} networks ---with a small length scale corresponding to high clustering coefficients and vice versa--- and obtained \OlavRev{the same} result, i.e. very similar dynamics for very different decay lengths (not shown).

\DemianRev{We stress that our procedure for the automatic generation of networks with similar bursting dynamics was not guaranteed to converge for such a wide range of clustering coefficients. Thus, the illustrative simulations of Fig.~\ref{fig:CCnets_all_look_alike} provide genuine evidence that the relation between network dynamics and network structural clustering is not trivially ``one-to-one''}, despite the fact that more clustered networks have been shown to have different cascading dynamics at the onset of a burst~\cite{Pajevic:2009p252}. 

\subsection*{Extraction of \DemianRev{directed functional} connectivity}

\DemianRev{We focus, first, on the reconstruction of simulated networks, taken from} the \textit{local} and \textit{non-local} ensembles described above. We compute their \DemianRev{directed functional} connectivity based on simulated calcium signals.

\subsubsection*{\DemianRev{Discrete differentiation}}

\DemianRev{Synthetic fluorescence time series are pre-processed only by simple discrete differentiation,
% as a rough attempt
such as to extract baseline modulations associated to potential firing. These differentiated signals are then used as input to any further analyses.}

\subsubsection*{\DemianRev{Generalized TE}}
We resort to a modified version of TE that includes two novel features (described in detail in the {\it Materials and Methods} section), namely the treatment of ``same bin interactions'' and the \DemianRev{ad hoc} selection of dynamical states.

The original formulation of TE was designed to detect the causal influence of events in the past with events at a later time. Practically, since calcium fluorescence is sampled at discrete times, standard TE evaluates the influence of events occurring in the time-bin~$t$ of events occurring in earlier time bins $t-1$, $t-2$, $\ldots$. By including \textit{same bin interactions} in TE estimation, we consider also potential causal interactions between events that occur within \OlavRev{the} same time-bin~$t$. This is important when dealing with experimental data of real neuronal cultures since the image acquisition rate is not sufficiently high to establish the temporal order of elementary spiking events.

On the other hand, the \textit{selection of dynamical states} is crucial to properly capture interactions between neurons which lead to different activity correlation patterns in different dynamical regimes. Both simulated and real neuronal cultures show indeed a dynamical switching between two distinct states (bursting and non-bursting) that can be separated and characterized by monitoring the average fluorescence amplitude and restricting the analysis only to recording sections in which this average fluorescence falls in a predetermined range. Selection of dynamical states is discussed in the next section.

\subsubsection*{\DemianRev{From TE evaluation to directed functional networks}}

% Cut, because already said above (Olav):
% The details of the actual reconstruction algorithm are fully described in the {\it Materials and Methods} section. We summarize here only its main steps. 

Once TE \DemianRev{functional} connectivity strengths have been calculated for every possible directed pair of nodes, a reconstructed network topology can be obtained by applying a threshold to the TE values at an arbitrary level. Only links whose TE value is above this threshold are retained in the reconstructed network topology. 

\subsubsection*{\DemianRev{Choosing a threshold is equivalent to choosing an average degree}}

\DemianRev{As a matter of fact, selecting a threshold for the inclusion of links corresponds to set the average degrees of the reconstructed network. As intuitive and as shown in Figure~S1A, a linear correlation exists between the number of links and the average degree. Because of this relation, an expectation about the probability of connection in the culture, and hence, its average degree, can directly be translated into a threshold number of links to include.} 

\DemianRev{Based on the aforementioned estimations of probability of connection and taken into account the different sizes of our (smaller) simulated network and of our (larger) experimental cultures, threshold values are roughly selected to include the \textit{top~10\% of links}, for reconstructions of simulated networks, and to include the \textit{top~5\% of links}, for reconstructions from actual biological recordings. These choices are such to lead, in both cases, to reconstructed networks with comparable probability of connection, as previously mentioned.}

\DemianRev{The (limited) impact of a ``wrong'' threshold selection on the inference of specific topologic features, like the clustering coefficient, will be discussed in later sections.}

\subsubsection*{\DemianRev{Extracted functional networks vs ground-truth structural networks}}

For the simulated data, the resulting connectivity matrix can be directly compared to the ground truth topology, and a standard \textit{Receiver-Operator Characteristic}~(ROC) analysis can be used to quantify the quality of reconstruction. ROC curves are generated by gradually moving a threshold level from the lowest to the highest TE value, and by plotting at each point the fraction of true positives as a function of the fraction of false positives. Examples of ROC curves are shown in Figs.~\ref{fig:state_dependency}--\ref{fig:reconstruction_dependencies} for different reconstruction conditions and network ensembles and are commented in the corresponding sections.

\subsubsection*{\DemianRev{Alternative functional connectivity metrics}}

\DemianRev{We compare directed functional connectivity based on TE with alternative networks based on (adapted) crosscorrelation (XC), Granger Causality (GC) or Mutual Information (MI) metrics. Detailed definitions of them are provided in the \textit{Materials and Methods} section. The only}
% required change
\OlavRev{difference} \DemianRev{with respect to the procedure just described is evaluating per each directed edge functional coupling scores based on XC, GC or MI, rather than on TE.}

\subsection*{Network reconstruction depends on the dynamical states}

Immediately prior to the onset of a burst the network is very excitable. In such a situation it is intuitive to consider that the \DemianRev{directed functional} connectivity can depart radically from the structural \DemianRev{excitatory} connectivity, because local events can potentially induce changes at very long ranges, due to collective synchronization, rather than to direct synaptic coupling. Conversely, in the relatively quiet inter-burst phases, a post-synaptic spike is likely to be influenced solely by the presynaptic firing history. Hence, the \DemianRev{directed functional} connectivity between neurons is intrinsically \textit{state dependent} (cfr. also \cite{Battaglia:2012eg}), a property that must be taken into account when reconstructing the connectivity.

\begin{figure}
    [!ht]
    \begin{center}
        \includegraphics[width=17.35cm]{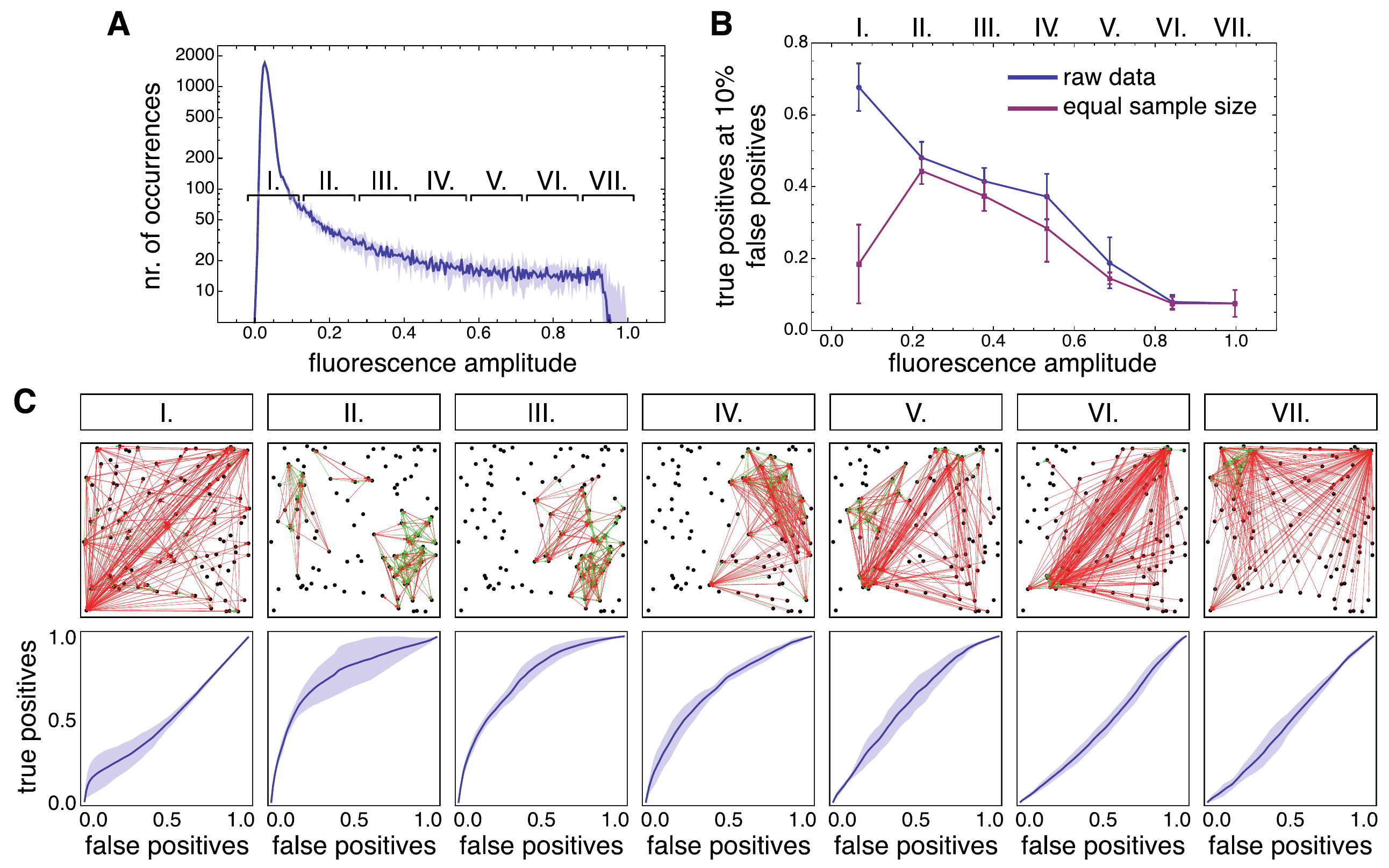}
    \end{center}
    \caption{} \label{fig:state_dependency}
\end{figure}

We illustrate here the state dependency of \DemianRev{directed functional} connectivity by generating a random network from the local clustering ensemble and by simulating its dynamics, including light scattering artifacts to obtain more realistic fluorescence signals. The resulting distribution of fluorescence amplitudes is divided into seven non-overlapping ranges \OlavRev{of equal width,}
% to explore a broad spectrum of dynamic regimes
each of them identified with a Roman numeral (Fig.~\ref{fig:state_dependency}A). Finally, TE \OlavRev{is computed} separately for each of these ranges, based on different corresponding subsets of data from the simulated recordings.

The quality of the reconstruction is \DemianRev{summarized} by the \textit{performance level}, \DemianRev{which, following an arbitrary convention, is measured as} the fraction of true positives at~10\% of false positives \DemianRev{read out of a complete ROC curve.}

We plot the performance level as a function of the average fluorescence amplitude in each interval, as shown in the blue line of Fig.~\ref{fig:state_dependency}B. The highest accuracy is achieved in the lowest fluorescence range, denoted by~I, \DemianRev{and reaches a remarkably elevated value of} approximately~70\% of true positives. The performance in the higher ranges~II to~IV decreases to a value around~45\%, to abruptly drop at range~V and above to a final plateau that corresponds to the~10\% performance of a random reconstruction (ranges~VI and~VII). 

Note that fluorescence values are not distributed homogeneously across ranges I-VII, as evidenced by the overall shape of the fluorescence distribution in Fig.~\ref{fig:state_dependency}A. For example, the lowest and highest ranges (I and~VII), differ \OlavRev{by} two orders of magnitude in the number of data points. To discriminate unequal-sampling effects from actual state-dependence phenomena, we studied the \OlavRev{performance level} using an equal number of data points in all ranges. Effectively, we restrict the number of data points available in each range to be equal to the number of samples in the highest range,~VII. The quality of such a reconstruction is shown as \OlavRev{the} red curve in Fig.~\ref{fig:state_dependency}B. The \OlavRev{performance level} is now generally lower, reflecting the reduced \OlavRev{number of time points which are included in the analysis}. 

Interestingly, the \DemianRev{``true''} peak of reconstruction quality is shifted to range~II, corresponding to fluorescence levels just above the Gaussian in the histogram of Fig.~\ref{fig:state_dependency}A. This range is \OlavRev{therefore} the most effective in terms of reconstruction \OlavRev{performance} for a given data sampling. 

For the ranges higher than~II, the reconstruction quality gradually decreases again to the~10\% performance of purely random choices in ranges~VI and~VII. \DemianRev{The effect of adopting a (shorter) equal sample size is particularly striking for range~I, which drops from the best performance level almost down to the baseline for random reconstruction. As a matter of fact, range~I is the one for which the shrinkage of sample length due to the constraint for uniform data sampling is most extreme (see later section on dependence of performance from sample size).}

The above analysis leads to a different \DemianRev{functional} network for each dynamical range studied. For the analysis with an equal number of data point per interval, the seven effective networks are drawn in Fig.~\ref{fig:state_dependency}C (for clarity only the top~10\% of links are shown). Each \DemianRev{functional} network is accompanied with the corresponding ROC curve. 

The lowest range~I corresponds to a regime in which spiking-related signals are buried in noise. Correspondingly, the associated \DemianRev{functional} connectivity is \DemianRev{practically} random, as indicated by a ROC curve close to the diagonal. \DemianRev{Nevertheless, information about structural topology is still conveyed in the activity associated to this regime and can be extracted through extensive sampling.}

\DemianRev{At the other extreme,} corresponding to the upper ranges~V to~VII ---associated to fully developed synchronous bursts--- \DemianRev{the functional connectivity has also a poor overlap with the underlying structural network. As addressed later in the \textit{Discussion} section, functional connectivity in regimes associated to bursting is characterized by the existence of hub nodes with an elevated degree of connection. The spatio-temporal organization of bursting can be described in terms of these functional connectivity hubs, since nodes within the neighborhood of a same functional hub experience a strongest mutual synchronization than arbitrary pair of nodes across the network (see \textit{Discussion} and also Figure~S2).}

The best agreement between \DemianRev{functional} and \DemianRev{excitatory} structural connectivity is clearly obtained for the inter-bursts regime associated with the middle range~II, and to a lesser degree in ranges~III and~IV, \DemianRev{corresponding to the early building-up of synchronous bursts}.

Overall, this study of state-dependent functional connectivity provides arguments to define the optimal dynamical regime for network reconstruction: The regime should include all data points whose average fluorescence across the population~$g_t$ is below a ``conditioning level''~$\tilde{g}$, located just on the right side of the Gaussian part of the histogram of the average fluorescence (see {\it Materials and Methods}). This selection excludes the regimes of highly synchronized activity (ranges~III to~VII) and keeps most of the data points for the analysis in order to achieve a good signal-to-noise ratio. \DemianRev{Thus, the inclusion of both ranges~I and~II combines the positive effects of correct state selection and of extensive sampling.}

\DemianRev{The state-dependency of functional connectivity is not limited to synthetic data. Very similar patterns of state-dependency are observed also in real data from neuronal cultures. In particular, in both simulated and real cultures, the functional connectivity associated to the development of bursts displays a stronger clustering level than in the inter-burst periods. An analysis of the topological properties of functional networks obtained from real data in different states (compared with synthetic data) is provided in Figure~S3.
% (more informations are provided in the caption).
In this same figure, sections of fluorescence time-series associated to different dynamical states are represented in different colors, for a better visualization of the correspondence between states and fluorescence values (for simplicity, only four fluorescence ranges are distinguished).}

\subsection*{\DemianRev{Analysis of two representative network reconstructions}}

Our generalized TE, conditioned to the proper dynamic range, enables the reconstruction of network topologies even in the presence of light scattering artifacts. For non-locally clustered topologies we obtain a \DemianRev{remarkably high} accuracy of up to~75\% of true positives at a cost of~10\% of false positives. An example of the reconstruction for such a network, with~$\text{CC}=0.5$, is shown in Fig.~\ref{fig:one_reconstruction_illustrated-CC}A. For locally-clustered topologies, accuracy typically reaches~60\% of true positives at a cost of~10\% of false positives, and an example for~$\lambda=0.25~\text{mm}$ is shown in Fig.~\ref{fig:one_reconstruction_illustrated-Lambda}A. 

In both topologies, we observe that for a low fraction of false positives detection (i.e. at high thresholds~$\Theta_{\text{TE}}$) the ROC curve displays a sharp rise, indicating a very reliable detection of the causally most efficient \DemianRev{excitatory} connections. A decrease in the slope, and therefore a rise in the detection of false positives and a larger confidence interval, is observed only at higher fractions of false positives. The confidence intervals are broader in the case of locally-clustered topologies because of the additional network-to-network variability that results from the placement of neurons (which is irrelevant for the generation of the non-locally clustered ensembles, see {\it Materials and Methods}).

\subsubsection*{\JordiRev{Non-local clustering ensemble}}
\begin{figure}
    [!ht]
    \begin{center}
        \includegraphics[width=12.35cm]{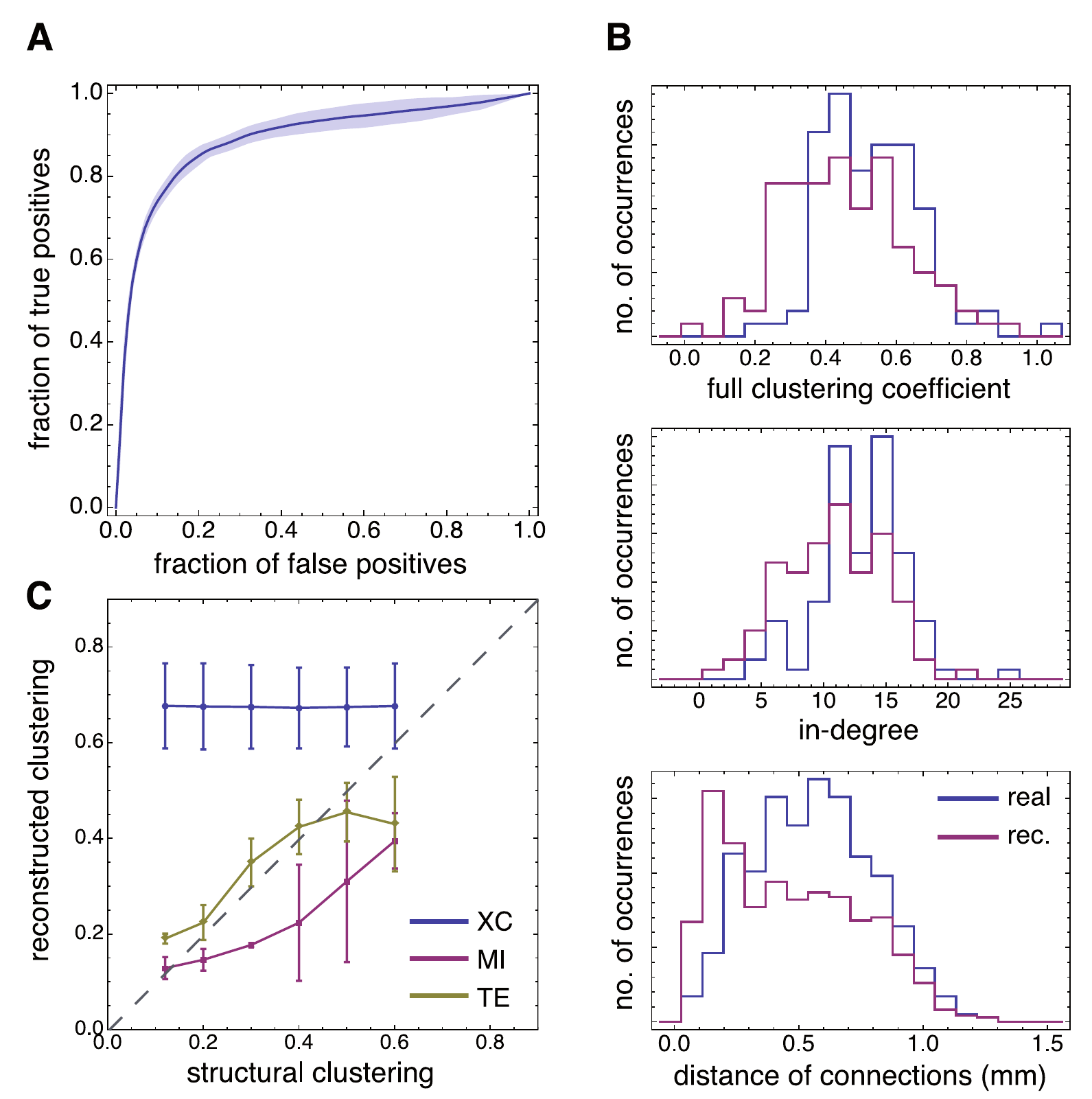}
    \end{center}
    \caption{} \label{fig:one_reconstruction_illustrated-CC}
\end{figure}

To address the reconstruction quality of the network topology, we focus first on the results for the non-local clustered ensemble. For a conditioning level which corresponds to the right hand side of the Gaussian in the fluorescence amplitude histogram ($\tilde{g}\simeq 0.2$), we consider three main network observables, namely the distributions of local clustering coefficients, in-degrees, and the distances of connections. As shown in Fig.~\ref{fig:one_reconstruction_illustrated-CC}B, we obtain a reconstructed network that reproduces well the ground truth properties, with similar mean values and distributions for all three observables considered. We observe, however, a small shift towards lower clustering indices (Fig.~\ref{fig:one_reconstruction_illustrated-CC}B, top panel) and especially towards lower average distances (bottom panel) for this highly clustered network.

Despite this underestimation bias for instances with high clustering, Fig.~\ref{fig:one_reconstruction_illustrated-CC}C shows the existence of a clear linear correlation between the real average clustering coefficient and the one of the topology reconstructed with generalized TE (Pearson's correlation coefficient of $r = 0.92$). Such linear relation allows, notably, a reliable discrimination between networks with different levels of clustering but very similar bursting dynamics. \DemianRev{Note that this linear relation between real and reconstructed clustering coefficient is robust against misestimation of the expected average degree, or, equivalently, of the number of links to include, as highlighted by Figure S1B.}

\JordiRev{TE-based reconstructions yield also estimates of the average distance of connection ---constant and not correlated with the clustering level--- with reasonable accuracy as shown in Supplementary Figure~S4A}. 

\subsubsection*{\JordiRev{Local clustering ensemble}}
%We reconstruct the average distance of connections also for the local clustering ensemble, in which this distance correlates with %the degree of clustering (Fig.~\ref{fig:one_reconstruction_illustrated-Lambda}). 

\begin{figure}
    [!ht]
    \begin{center}
      \includegraphics[width=17.35cm]{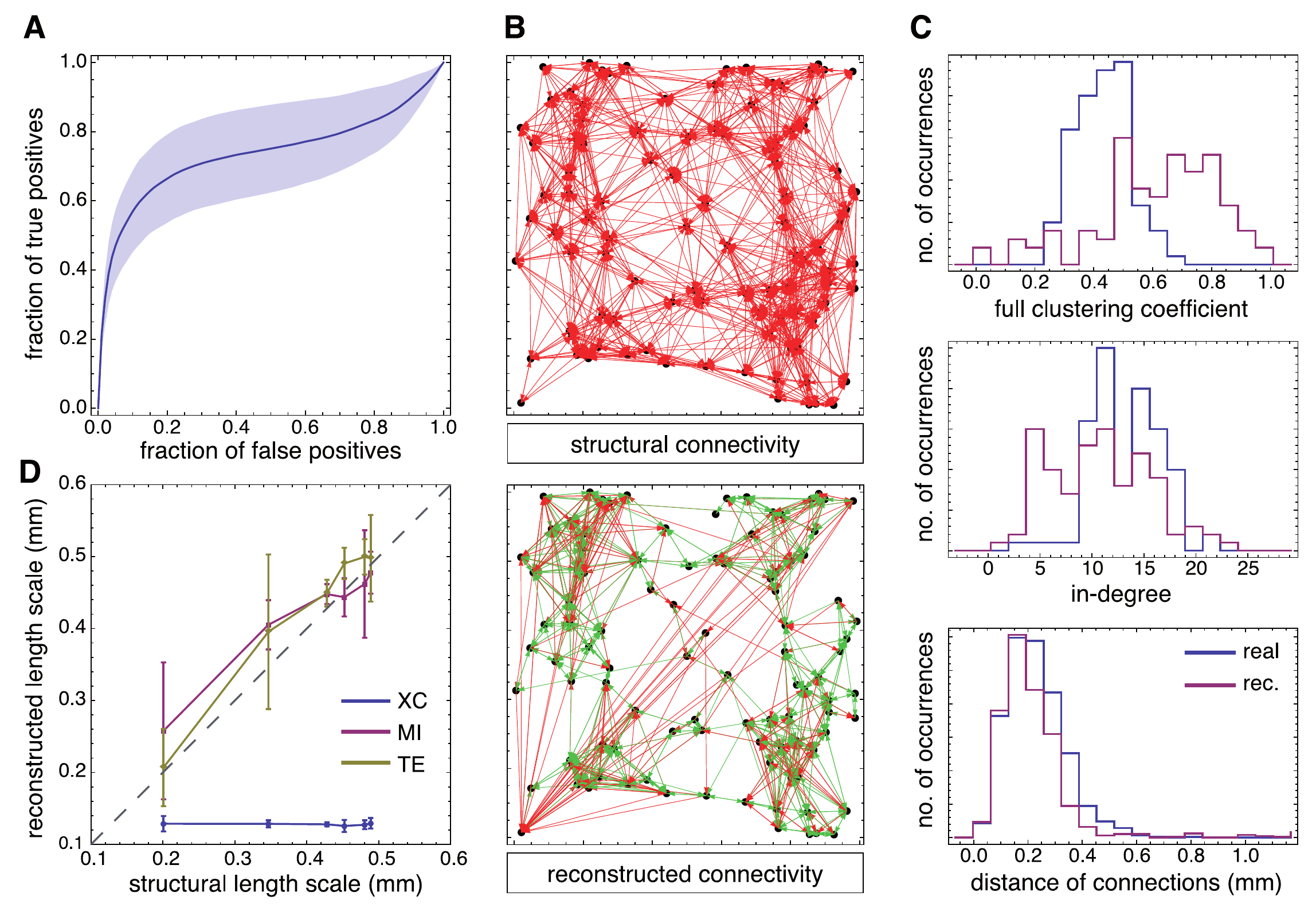}
    \end{center}
    \caption{} \label{fig:one_reconstruction_illustrated-Lambda}
\end{figure}

For this ensemble \JordiRev{(see Fig.~\ref{fig:one_reconstruction_illustrated-Lambda}), the} quality of reconstruction can be assessed even visually, \OlavRev{by plotting the network} graph of reconstructed connections, due to the distance-dependency of the connections. In Fig.~\ref{fig:one_reconstruction_illustrated-Lambda}B we compared the structural network (top panel) with the reconstructed one (bottom panel), \DemianRev{obtained by including as links only edges corresponding to the top 10\% of TE values. This corresponds here to} \OlavRev{about 600 true positives ($\sim 50\%$ of all possible true positives, and plotted in green) and about 400 false positives ($\sim 5\%$ of all possible false positives, plotted in red).}
% In the bottom diagram, edges colored in green correspond to correctly reconstructed edges.
The statistical properties of the structural and reconstructed networks are shown in Fig.~\ref{fig:one_reconstruction_illustrated-Lambda}C.
% for the distribution of local clustering coefficients, degree distribution, and distance of connections.
Again, reconstructed network properties correlate with real properties. The reconstructed distribution of connection distances displays a reduced right-tail compared to the real one. A tendency to estimate a more local connectivity is evident also from a marked overestimation of local clustering coefficients. We attribute such a mismatch to light scattering artifacts that increase local correlations in a spatial region matching the length scale of real structural connections. This is confirmed by the fact that the length scale is correctly inferred in simulations without the light scattering artifact (not shown).

Note, that there is \OlavRev{again a very} good linear correlation (Pearson's correlation of $r = 0.97$) between the actual and reconstructed (spatial) average connection length, as shown in Fig.~\ref{fig:one_reconstruction_illustrated-Lambda}D. Similarly, the reconstructed average clustering coefficient is linearly correlated with the ground truth one ($r = 0.98$), as shown in Supplementary Figure~S4B. 

\subsubsection*{\JordiRev{Sensitivity to reconstruction approaches}}

\DemianRev{Overall, TE of Markov order~$k=2$ (i.e. taking into account multiple time scales of interaction, see eq. 9 in Methods section) achieved a performance} \OlavRev{level} {ranging between~40\% and~80\% at a level of~10\% of false positives, for any clustering type and level.}

\DemianRev{In Fig.~\ref{fig:one_reconstruction_illustrated-CC}C and Fig.~\ref{fig:one_reconstruction_illustrated-Lambda}D} we compare the performance of generalized TE with other reconstruction strategies, \DemianRev{respectively for the non-local and for the local clustering ensemble.} We observe then that MI-based reconstructions yield linear correlations between real and reconstructed clustering coefficient and length scales as well. \DemianRev{For the adopted optimal conditioning level, MI can actually out-perform generalized TE (cfr. Supplementary Figure~S4), due probably to the smaller sample size required for its estimation.} On the contrary, XC-based reconstructions fail \OlavRev{in all cases to reproduce} these linear correlations, \OlavRev{yielding} a constant value independently from the ground-truth values. For the non-local clustering ensemble, it distinctly over-estimates average clustering level; for the local clustering ensemble, it severely underestimates the average length of connections. Therefore, in XC-based reconstructions, all information on the actual degree of clustering in the network is lost and a high clustering level is invariantly inferred.

GC-based reconstructions display the same error syndrome (not shown), which indicates that capturing non-linear correlations in neural activity ---as MI and TE can do, but XC and GC cannot--- is crucial for the inference of the clustering level.

\DemianRev{We would like to remind that XC-, GC- and MI-based methods, analogously to the generalized TE approach, include, as generalized TE, the possibility of ``same-bin interactions'' (zero-lag). \DemianRev{Furthermore,} \JordiRev{we have modified them to include} \DemianRev{as well} \JordiRev{optimal conditioning in order to make the comparison between different methods fair.} The forthcoming section gives more details about comparison of performance for different conditions and methods.}

\subsection*{\DemianRev{Performance comparison:} Role of topology, dynamics and light scattering}

The performance level (fraction of true positives for 10\% of false positives, \OlavRev{denoted by} \JordiRev{TP$_{10\%}$}) provides a measure of the quality of the reconstruction, and allows the comparison of different methods for different network topologies, conditioning levels, and external artifacts (i.e. presence or absence of simulated light scattering). We test linear methods, XC and GC (of order~2; the performance of GC of order~1 is very similar and not shown), and non-linear methods, namely MI and TE (of Markov orders~1 and~2; \DemianRev{see \textit{Materials and Methods} for details}). XC and MI are correlation measures, while GC and TE are causality measures. Note that, for each of these methods, we account for state dependency of \DemianRev{functional} connectivity, performing state separation as described in the {\it Materials and Methods} section.

\begin{figure}
    [!ht]
    \begin{center}
        \includegraphics[width=17.35cm]{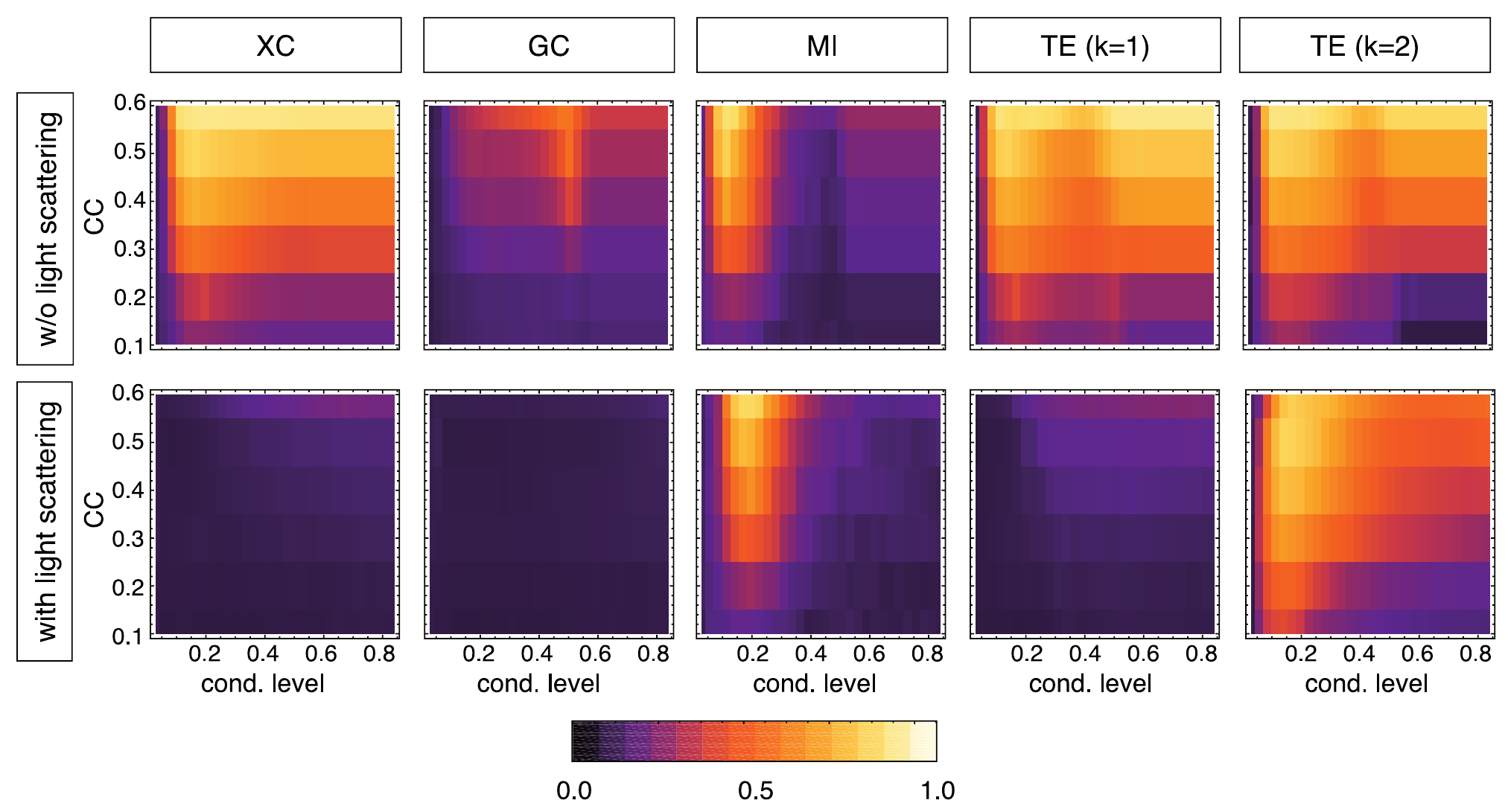}
    \end{center}
    \caption{} \label{fig:2D-scans}
\end{figure}

% We focus here on the non-local clustering ensemble and show the results of our comparisons in Fig.~\ref{fig:2D-scans}. Supplementary Figure~S5 reports analogous results for the local clustering ensemble.
\OlavRev{In the case of the non-local clustering ensemble and} without light scattering (Fig.~\ref{fig:2D-scans}, top row), even a linear method such as XC achieves a good reconstruction. This success indicates an overlap between communities of higher synchrony in the calcium fluorescence, associated to stronger activity correlations, and the underlying structural connectivity, especially for higher full clustering indices.

GC-based reconstructions have an overall worse quality, due to the inadequacy of a linear model for the prediction of our highly nonlinear network dynamics, but they show similarly improved performance for higher~$\text{CC}$.

In a band centered around a shared optimal conditioning level $\tilde{g}\simeq 0.2$, both MI and generalized TE show a robust performance across all clustering indices. This value is similar to the upper bound of the range~II depicted in Fig.~\ref{fig:state_dependency}A, i.e. it lies at the interface between the bursting and silent dynamical regimes. In particular for TE and in the case of low clustering indices (which leads to networks closer to random graphs), conditioning greatly improves reconstruction quality. At higher clustering indices the decay in performance is only moderate for conditioning levels above the optimal value, indicating an overlap between the \DemianRev{functional} connectivities in the bursting and silent regimes. \DemianRev{Note, on the contrary, that the performance of MI rapidly decreases if a non-optimal conditioning level is assumed.}

The introduction of light scattering causes a dramatic drop in performance of the two linear methods (XC and GC), and even of \DemianRev{MI and} TE with Markov order $k=1$. The performance of TE at Markov order~2 also deteriorates, but is still significantly above the random reconstruction baseline in a broad region of parameters. Interestingly, for the optimal conditioning level $\tilde{g}\simeq 0.2$ the performance of the TE for $k=2$ does not fall below TP$_{10\%}\sim 40\%$ for any clustering level or~$\lambda$ value. It is precisely in this optimal conditioning range that we obtain the linear relations between reconstructed and structural clustering coefficients, for both the non-local and the local clustering ensembles.

A similar trend is obtained when varying the length scale~$\lambda$ in the local ensembles (see Supplementary Figure~S5). For very local clustering and without light scattering, both XC and TE achieve performance levels up to~80\%. The introduction of light scattering, however, reduces the performance of all measures except \DemianRev{for MI (but only in the narrow optimal conditioning range) and} for TE of higher Markov orders \DemianRev{(robust against non-optimal selection of conditioning level)}. Overall, the performance of the reconstruction for the local clustering ensembles is lower than for the non-locally clustered ensembles. \DemianRev{This is true, incidentally, also in absence of light scattering, since networks sampled from this ensemble,  as soon as the length scale is \JordiRev{sufficiently long}, tend to be very similar to purely random topologies (of the Erd\"os-R\'enyi type, see e.g. \cite{Newman:2001un}), for which performance is generally poorer (cfr. top row of Figure~\ref{fig:2D-scans}, for weak clustering levels).}

\subsection*{\DemianRev{Contributions to the performance of generalized TE}}

\begin{figure}
    [!ht]
    \begin{center}
        \includegraphics[width=12.35cm]{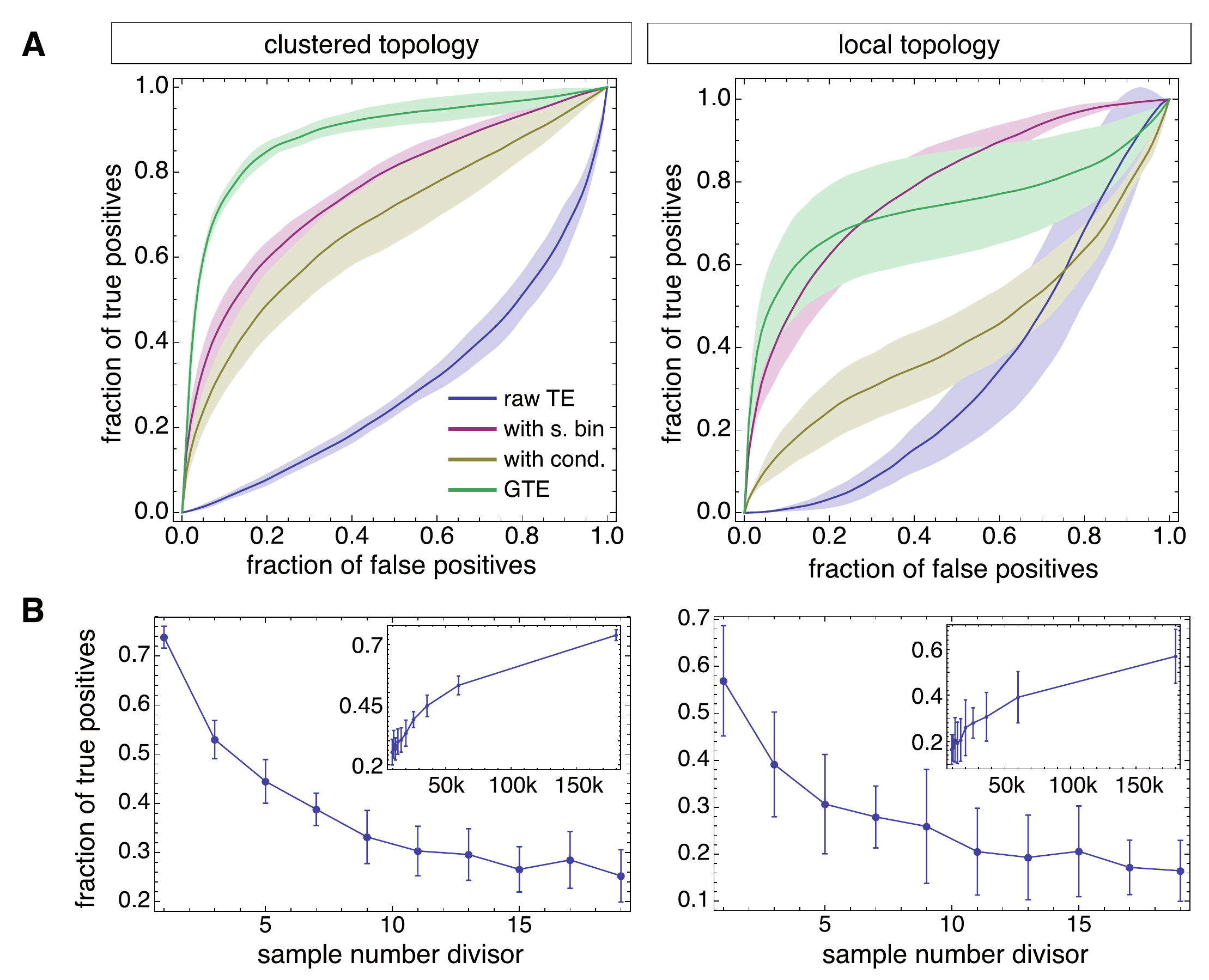}
    \end{center}
    \caption{} \label{fig:reconstruction_dependencies}
\end{figure}

Our new TE method significantly improves the reconstruction performance compared to the original TE formulation \JordiRev{[Eq.~(\ref{eq:TE-def})]}. As shown in Fig.~\ref{fig:reconstruction_dependencies}A for both the local and the non-local clustered networks, reconstruction with the original TE formulation \JordiRev{(Fig.~\ref{fig:reconstruction_dependencies}A, blue line) yields worse results than a random reconstruction}\DemianRev{,  as indicated by the corresponding ROC curves falling below the diagonal}. Such a poor success is due in large part to ``misinterpreted'' delayed interactions. Indeed, by taking into account same bin interactions, a boost in performance is observed (red line). Fig.~\ref{fig:reconstruction_dependencies}A also shows that an additional leap in performance is obtained when the analysis is \textit{conditioned} (i.e. restricted) to a particular dynamical state of the network, increasing reconstruction quality by~20\% (yellow line in Fig.~\ref{fig:reconstruction_dependencies}A). The determination of the optimal conditioning level is discussed later and takes into account the considerations introduced \OlavRev{above} (cfr.
Fig.~\ref{fig:state_dependency}). 

\DemianRev{Note that the introduction of same bin interactions alone (red color curves) or conditioning on the dynamical state of network alone (yellow color curves) already brings the performance to a level well superior to random performance. However,} at least for our simulated calcium-fluorescence time series, a remarkable boost in performance is obtained \DemianRev{only when same-bin interactions inclusion and optimal conditioning are combined together (green color curves). Although, in principle, conditioning is enough to select indirectly a proper dynamical regime, the poor time-resolution of the analyzed signals (constrained not only by the frame-rate of acquisition but also intrinsically by the kinetics of the dissociation reaction of the calcium-sensitive dye \cite{Eberhard:1989ve}) requires as well the potential consideration of causally-linked events occurring in a same time-bin.}

\OlavRev{A different way to represent reconstruction performance are ``Positive Precision Curves'' (as introduced in~\cite{Garofalo:2009jm} and described in the \textit{Materials and Methods} section) by plotting, at a given number of reconstructed links, the ``true-false ratio'' (TPR), emphasizing the probability that a reconstructed link is present in the ground truth topology (true positive). For the same networks and reconstruction as above, we plot the PPCs in Fig.~S6 (for the ROC curves see Fig.~\ref{fig:reconstruction_dependencies}A). Over a wide range of the number of reconstructed links (TFS), the PPC displays positive values of the TFR, indicative of a majority of true positives over false positives. For both the locally and non-locally clustered networks, the PPC reaches a maximum value of the TPR about~0.5} \DemianRev{and remains positive up to about~18\% of included links for the clustered topology, or up to~12\% in the case of the local topology.}

\subsection*{\DemianRev{Recording length affects performance}}

In Fig.~\ref{fig:reconstruction_dependencies}B, we analyze the performance of our algorithm against changes of the sample size. Starting from simulated recordings lasting 1h of real time \OlavRev{(corresponding to about 360 bursting events)} and with a full sample number of of $S_{\text{1h}}$, we trimmed these recordings producing shorter fluorescence time series with ~$S'=S_{\text{1h}}/s$ samples, with $s$ being a divisor of the sample size. For both network topology ensembles, we found that \OlavRev{a reduction in the number of samples by a factor of two (corresponding to 30 minutes or about 180 burst) still yields a performance level of $\sim 70\%$. By further reducing the sample size, we reach a plateau with a quality of $\sim 30\%$ for about 40 bursts (corresponding to 6 minutes).}

All the experiments analyzed in this work are carried out with a duration between~30 and 60~minutes. Since conditioning, needed to achieve high performance, requires \DemianRev{one} to ignore a conspicuous fraction of the recorded data, we expect long recordings to be necessary for a good reconstruction, albeit the fact that it is possible to increase the signal-to-noise ratio by increasing the intensity of the fluorescent light. However, \JordiRev{the latter manipulation has negative implications for the health of neurons due to photo-damage, limiting our experimental recordings to a maximum of 2 hours.}

\subsection*{Analysis of biological recordings}

We apply our analysis to actual recordings from \emph{in vitro} networks derived from cortical neurons (see \emph{Methods}). To simplify the network reconstruction problem, experiments are carried out with blocked inhibitory GABA-ergic transmission, so that the network activity is driven solely by excitatory connections. This is consistent with previously discussed simulations, in which only excitatory neurons were included.

We consider in Fig.~\ref{fig:real_data} a network reconstruction based on a 60~minutes recording of the activity of a mature culture, at day \emph{in vitro} (DIV)~12, in which $N=1720$ active neurons were simultaneously imaged. A fully analogous network reconstruction for a second, younger dataset at DIV~9 is presented in Supplementary Figure~S7. \JordiRev{In general, fluorescence data was not affected neither by photo-bleaching nor by photo-damage during this time, as proved by the stability of the average fluorescence signal shown in the Supplementary Figure~S8A}. 

The probability distribution of the average fluorescence signal is computed in the same way as for the simulated data. Neuronal dynamics and the calcium fluorescence display the same bursting dynamics well captured by the simulations, leading to \JordiRev{a similar fluorescence distribution (Fig.~\ref{fig:raw_data_vs_simulations}D)}. Thanks to this similarity we can make use of the intuition developed for synthetic data to estimate an adequate conditioning level. We select therefore a conditioning level such as to exclude the right-tail of high fluorescence associated to to fully-developed bursting transient regimes. We have verified, however, that the main qualitative topological features of the reconstructed network are left unchanged when varying the conditioning level in a range centered on our ``optimal'' selection. More details on conditioning level selection are given in the {\it Materials and Methods} section.

Reconstruction analysis is carried out for the entire population of imaged neurons. We analyze a network defined by the top~5\% of TE-ranked links, \DemianRev{as discussed in a previous section. Such choice} leads to an average \JordiRev{in--}degree of about~100, compatible with average degrees reported previously for neuronal cultures of corresponding age (DIV) and density~\cite{Soriano:2008p747,Jacobi:2009gn}.

\subsubsection*{\DemianRev{Comparison with randomized networks}}

The ground truth \DemianRev{excitatory} connectivity is obviously not known for real recordings and performance cannot be assessed by means of an ROC analysis. However, we can compare the obtained reconstruction to randomized variations to identify non-trivial topological features of the reconstructed network. We perform two kinds of randomizations: In a first one, randomization is full and only the number of network edges is preserved. Comparison with such \emph{fully} randomized networks detects significant deviations of the reconstructed network from an ensemble of random graphs in which the degree follows the same prescribed Poisson distribution for each node (Erd\"os-R\'enyi ensemble, see e.g. \cite{Newman:2001un}). In a second randomization, both the total edge number and the precise out-degrees of each node are preserved. The comparison of the reconstructed network with such a \emph{partially} randomized ensemble detects local patterns of correlations between in- and out-degrees ---including, notably, clustering--- which do not arise just in virtue of a specific distribution of out-degrees. Comparison with partial randomizations is particularly important when skewed distributions of degrees are expected.

\subsubsection*{\DemianRev{Topology retrieved by TE}}

\begin{figure}
    [!ht]
    \begin{center}
        \includegraphics[width=17.35cm]{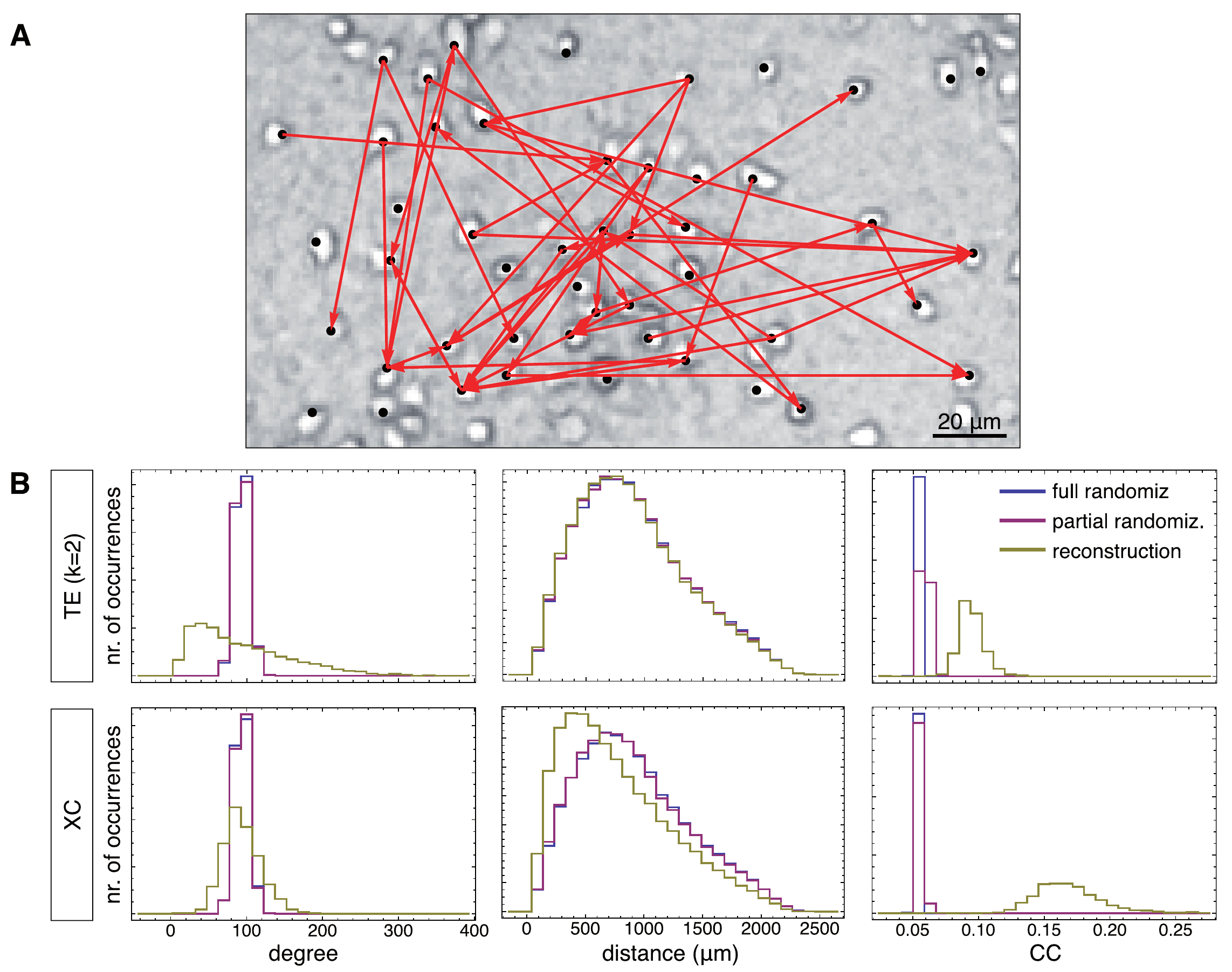}
    \end{center}
    \caption{} \label{fig:real_data}
\end{figure}

Our analysis shows that the resulting reconstructed topology is characterized by markedly non-local structures, as visible in the portion of the reconstructed network in Fig.~\ref{fig:real_data}A. Distributions of degree, distance of connection and local clustering coefficients inferred by TE are shown in the top row of Fig.~\ref{fig:real_data}B (yellow histograms). The degree distribution is characteristically broadened and distinctly right-skewed, deviating from the Poisson distribution associated to Erd\"os-R\'enyi random graphs (the histogram for fully randomized networks shown in blue). Note that, for partial randomizations (histograms shown in red), we have randomized the out-degree of each node but plotted here the resulting in-degree distribution (the distribution of out-degrees would be, by construction, unchanged).

While the distribution of connection distances matches the one of randomized networks, TE detects clustering at a level which is moderate ($\text{CC}\simeq 0.09$) but significantly larger than for random networks. Note that this larger clustering cannot be ascribed to the broadened degree distribution since both full and partial (red histograms) randomizations lead to consistently smaller clustering levels ($\mathrm{CC}\simeq 0.05$).

\subsubsection*{\DemianRev{Topology retrieved by XC}}

By analyzing a network reconstruction based on cross-correlation (XC), we find differences to TE (bottom row of Fig.~\ref{fig:real_data}B). In particular, XC infers a distribution of distances markedly more local than for full and partially randomized network instances and, correspondingly, a markedly higher clustering ($\mathrm{CC}\simeq 0.17$). Distribution of degrees inferred by XC is on the contrary random-like.

As a matter of fact, remarkably similar patterns of discrepancy between reconstruction results based on TE and on XC are also robustly present in synthetic data. Synthetic data analyses consistently show a superior performance of TE compared to XC. Furthermore, these analyses identify a tendency of XC to infer an artificially too local and too clustered connectivity. Therefore, we believe that the topology of the neuronal culture inferred by XC is not reliable, and is biased by the aforementioned systematic drifts.

\section*{Discussion}

\subsection*{\DemianRev{Relation to state of the art}}

We have introduced a novel extension of Transfer~Entropy, an information theoretical measure, and applied it to infer \DemianRev{excitatory} connectivity of neuronal cultures \emph{in vitro}. \DemianRev{Other studies have previously applied TE (or a generalization of TE) to the reconstruction of the topology of cultured networks \cite{Garofalo:2009jm, Ito:2011wc}. However, our study introduces and discusses important novel aspects, relevant for applications. }

\subsubsection*{\DemianRev{Model independency}}

Our algorithm is model-independent and can be used virtually without modifications even for the reconstruction based on spike trains or voltage traces. This is important, since massive datasets with modalities beyond calcium fluorescence imaging might become available in a near future, thanks to progresses in connectomics research. This model-independence, as previously mentioned, is also important to avoid potential artifacts due to a wrong choice of model for neuronal firing or for network topology. Therefore, it constitutes a major advantage with respect to regression methods or even more elaborated Bayesian approaches, as the one considered in~\cite{Mishchencko:2009p4301}. Both regression and Bayesian techniques indeed assume specific models of calcium fluorescence and neuronal firing dynamics, either explicitly (in the case of the Bayesian framework) or implicitly (assuming a linear dynamical model in the case, e.g., of XC or GC). \DemianRev{Note that we use here dynamical network models to benchmark our reconstruction quality. However, our method still remains model-free, because knowledge about these models is not required for reconstruction.}

\subsubsection*{\DemianRev{No need for spike times}}

Competitor approaches \DemianRev{used~\cite{Garofalo:2009jm} or} put emphasis on the need of reconstructing exact spike times~\cite{Mishchencko:2009p4301} with sophisticated deconvolution techniques~\cite{Vogelstein:2009p3029}, as a preprocessing step before actual topology reconstruction. 

As we have shown here, \DemianRev{acquiring such difficult-to-access information} is unnecessary for our method, \DemianRev{which performs} efficiently even for slow calcium fluorescence acquisition rate and \DemianRev{operates} directly on imaging time series. \DemianRev{This is a crucial feature for applications to noisy data, which remains useful even when ---as for the data analyzed in this study--- the signal-to-noise ratio is sufficiently good to allow sometimes the isolation of individual firing events (cfr. Fig.~\ref{fig:raw_data_vs_simulations}B).}

\subsubsection*{\DemianRev{Robustness against bursting}}

% Another important aspect of our study is the fact that 
We optimized our algorithm to infer \DemianRev{excitatory} connectivity based on time series of calcium fluorescence with a complex nonlinear dynamics, capturing the irregular bursting and the corresponding time-dependent degree of synchronization observed in cultured networks \emph{in vitro}. To our knowledge, no previous study about algorithmic connectivity reconstruction has tackled with simulated dynamics reaching this level of realism. We have here identified a simple and conceptually elegant mean-field solution to the problem of switching between bursting and non-bursting states, based just on conditioning with respect to the average level of fluorescence from the whole culture.

A feature of our model network dynamics, and \OlavRev{one} that is crucial to \DemianRev{reproduce temporally irregular network bursting}, is the inclusion of short-term depressing synapses. Remarkably, other studies~\cite{Ito:2011wc}, which have modeled explicitly more complex forms of spike-time dependent synaptic plasticity, neglect completely this short-term plasticity, failing correspondingly to generate a realistic model of spontaneous activity of an \emph{in vitro} culture. \DemianRev{Yet, our network model remains very simplified, although the use of networks of integrate-and-fire neurons to generate surrogate data is widespread \cite{Garofalo:2009jm, Mishchencko:2009p4301, Ito:2011wc}. Several features of real cultured neurons are not explicitly included, like heterogeneity in synaptic conductances and time-constants, slow NMDA excitatory currents or distance-dependent axonal delays. However, time-series from more complex models could be used to benchmark our algorithm, without need of introducing any change into it, due to its model-free nature.}

\subsubsection*{\DemianRev{Robustness against light scattering}}

We have found that, among the tested methods, only \JordiRev{generalized} TE of at least Markov order~$k=2$ with a proper conditioning allows to distinguish random from clustered topologies and local from long-range connectivities in a reliable manner, in the presence of light scattering artifacts. These artifacts \DemianRev{indeed} lead to the inference of spurious interactions between the calcium signal of two nodes, reducing the performance of linear causality measures \DemianRev{like XC or GC} to a random level.
% even if they are weak in amplitude.
Note that this is very likely a similar effect as in~\cite{Vicente:2011p6560}, where reconstruction with TE is still possible despite cross-talk between EEG electrodes.

\subsubsection*{\OlavRev{Low} computational complexity}

\DemianRev{Finally, we note that } our algorithm is computationally simple and relatively efficient. \DemianRev{Preprocessing of time series is a simple discrete differentiation. State selection is achieved via conditioning data on a range, which requires only to read once the input time series and compare them with a threshold. The inference procedure itself is not iterative (unlike in \cite{Pajevic:2009p252, Mishchencko:2009p4301}) and evaluation of TE is done through simple ``plug-in'' estimation, which is fast (unlike elementary steps in, e.g., \cite{Mishchencko:2009p4301}).} The principal determinant of computational complexity is therefore the growing number of putative links (which grow as $O(N^2)$) \DemianRev{and, hence, of TE scores that must be computed.}

For the recording duration of one hour considered in this work, we typically found a computation time of approximately $T_{\text{comp}} \approx 60ms * N^2$ including pre-processing on a 2.67~GHz Intel Xeon processor. Reconstructions of networks with 100 nodes required roughly ten minutes; with 1000 nodes roughly half a day.
% Networks of 10000 nodes might be reconstructed within months.
Note however that as the computation of TE for two distinct links is (after pre-processing and conditioning) computationally independent, it can be easily parallelized, reducing the computation time by a factor equal to the number of CPUs.

%
%\OlavRev{(Cut next paragraph?)}
%Furthermore, the computational demand is considerably decreased when the reconstruction of a whole network is not required. %For instance, to infer the existence of a specific link, its associated TE score might be tested statistically against a distribution of %TE values sampled only over a large but limited number of neuronal pairs.

\subsection*{Good reconstruction of topological features}

\subsubsection*{\DemianRev{Good reconstruction of clustering}}

Based on synthetic time series of calcium fluorescence, we have studied the relation between ground truth topological features and their reconstructed counterparts. By restricting \DemianRev{directed functional} connectivity estimation to a proper dynamical regime through conditioning, we found strong linear correlations between real and reconstructed topological properties, \OlavRev{both for} the average Euclidean distance of connections and the clustering coefficient (CC). Note that deviations from this linear relationship at higher CC values, visible in Fig.~\ref{fig:one_reconstruction_illustrated-CC}C, are \OlavRev{in part} due to the fact that we have used a constant conditioning level for all reconstructions, while the optimal conditioning level increases slightly for high CCs.

\DemianRev{Clustering coefficient is the most widespread measure of higher order topological features, going beyond characterization of neighbors in a network, but analyzing correlation between local neighborhoods of different nodes.} We have \DemianRev{adopted} the so-called ``full'' clustering index to measure clustering in our directed networks. However, several other definitions of clustering coefficient for directed network exist, emphasizing the contribution to the clustering phenomenon of different topological motifs such as cycles or ``middleman'' loops~\cite{Fagiolo:2007p53}. We have checked that the linear relationship between real and reconstructed clustering indices hold for all the clustering index types defined in~\cite{Fagiolo:2007p53}, with an almost identical degree of correlation. This \DemianRev{hints}
%, although indirectly, to}
\OlavRev{at} a good capacity of our algorithm to reconstruct different classes of graph topology motifs. 

\subsubsection*{\DemianRev{Good reconstruction of connectivity motifs}}

\DemianRev{We have focused on the performance of our algorithm in reconstructing specific topological motifs involving}
% few nodes at a time.
\OlavRev{more than two nodes.} \DemianRev{Considering for instance the network whose reconstruction is illustrated in Fig.~\ref{fig:one_reconstruction_illustrated-Lambda}, we identified in the ground truth topology of this network all occurrences of \textit{shared source motifs}, i.e. motifs in which a node~$A$ was connected to a node $B$ ($A \rightarrow B$) and to a node $C$ ($A \rightarrow C$) but in which there was no direct connection between $B$ and $C$} \OlavRev{in the reconstructed as well as the ground truth topology.} {Such a motif might be harmful for reconstruction, because the existence of a shared input~$A$ might be mistaken for a reciprocal causal interaction between nodes $B$ and $C$. Nevertheless, only in a minority of cases (\JordiRev{about 20\%}), spurious links $B\rightarrow C$ or $C\rightarrow B$ where erroneously included in our reconstruction. Equivalently, in the case of embedded chains, i.e. motifs in which a node~$A$ is connected to~$B$, connected on its turn to~$C$, without a direct link from $A$ to $C$ ($A \rightarrow B \rightarrow C$), the presence of a link $A \rightarrow C$ (reflecting potentially indirect causation, mistaken for direct) was erroneously inferred \JordiRev{only in, once again, about 20\% of cases.}} Note that other studies have investigated the performance of various metrics in the reconstruction of specific small network motifs \cite{Gourevitch:2006cn, Quinn:2011jt}, but it is not clear that the efficiency of reconstruction quantified for such small networks continues to hold when these motifs are embedded in larger networks with hundreds of nodes or more.

\DemianRev{Good reconstruction of higher-order topological features is important since these features (like e.g. the tendency of existing links to form chains) are known to affect the synchronizability of networks, as stressed by a recent study \cite{Zhao:2011cq}. Our algorithm was not specifically optimized to infer higher order structures. Analyses based on $k$-ples rather than on pairs of nodes might lead to a better description of higher order connectivity structures, at a price, however, of a more severe sampling problem. In this sense, conditioning transition probabilities for a specific edge to the ongoing mean-field activity, constitute already a first compromise, allowing to extend the analysis beyond a mere pairwise approach.}

\subsubsection*{\DemianRev{Overestimation of bidirectional links}}

\DemianRev{In \cite{Zhao:2011cq}, the role played by the fraction of bidirectional links (i.e. situations in which there is a connection $A\rightarrow B$, but also a connection $B\rightarrow A$) was also explored.} 

\JordiRev{In our case,}\DemianRev{ approximately 60\% of the bidirectional links present in the ground truth topology were reconstructed as bidirectional in the reconstruction of Fig.~\ref{fig:one_reconstruction_illustrated-CC}. However, this reconstruction exhibited as well a severe overestimation of the number of bidirectional links. Unidirectional in the ground-truth topology were more difficult to detect, such as only 4\% of them were ranked among the top~10\% of TE scores. Furthermore, when such unidirectional links $A \rightarrow B$ were reconstructed, the presence of a link $B\rightarrow A$ was also in most of the cases (85\%) spuriously inferred, i.e. included unidirectional links were often mistaken for bidirectional links.} 

\DemianRev{It is likely that an improvement in the inference of directionality might be achieved by increasing the time-resolution of the recordings beyond the present limits of calcium imaging techniques.}

\subsubsection*{\DemianRev{Good reconstruction of connectivity length scale}}

Finally, we have also analyzed the distance dependence of the reconstructed probability of connection in the system. We have found a good agreement between real and reconstructed average length scale of connections, albeit we did find more local connections than present in the ground truth topology (see for example Fig.~\ref{fig:one_reconstruction_illustrated-CC}B, bottom panel). This is an artifact due to light scattering, as it is suggested by the underestimated peak of the reconstructed connection distance histogram, which matches the characteristic length scale of simulated light scattering $\lambda_{\text{sc}}$.

\subsection*{\DemianRev{Functional connectivity hubs}}

\DemianRev{As shown in Fig.~\ref{fig:state_dependency} for simulated data and in Supplementary Fig.~S3 for actual culture data, epochs of synchronous bursting are associated to functional connectivity with a stronger degree of clustering and a weaker overlap with the underlying structural topology. This feature of functional connectivity is tightly related to the spatio-temporal organization of the synchronous bursts.}

\DemianRev{In figure S2A, we highlight the position of selected nodes (of the simulated network considered in Fig.~\ref{fig:state_dependency}), characterized by an above-average in-degree of functional connectivity (see \textit{Materials and Methods} for a detailed definition in terms of TE scores). We denote these nodes ---different in general for each of the dynamic regimes numbered from I to VII--- as (state-dependent) \textit{functional connectivity hubs}.} 

\DemianRev{Given a specific hub, we can then define the community of its first neighbors in the corresponding functional network. Consistently across all dynamic ranges (but the noise-dominated range I) we find that synchronization within each of these functional communities is significantly stronger than between the communities centered on different hubs (($p < 0.01$, Mann-Whitney test, see \textit{Materials and Methods}). The results of this comparison are reported in Figure~S2B, showing particularly marked excess synchronization during burst build-up (ranges II, III and IV) or just prior to burst waning in the largest-size bursts (VII).} Therefore, functional connectivity hubs reflect foci of enhanced local bursting synchrony. 

\DemianRev{Other studies (in brain slices) reported evidence of functional connectivity hubs, whose direct stimulation elicited a strong synchronous activation \cite{Morgan:2008hg, Bonifazi:2009in}. In \cite{Morgan:2008hg}, the functional hubs were also structural hubs. In the case of our networks, however only functional hubs associated to ranges II and III have an in-degree (and out-degree) larger than average as in \cite{Morgan:2008hg}. In the other dynamic ranges, this tight correspondence between structural and functional hubs does not hold anymore. Nevertheless, in all dynamic ranges (but range I), we find that pairs of functional hubs have an approximately three-times larger chance of being structurally connected than pairs of arbitrarily selected nodes (not shown). }

\DemianRev{The timing of firing of these strong-synchrony communities is analyzed in Figure S2C. There we show that the average time of bursting of functional communities for different dynamic ranges is shifted relatively to the average bursting time over the entire network (details of the estimation are provided in \textit{Materials and Methods}).} This temporal shift is negative for the ranges~II and~III (indicating that functional hubs and related communities fire on average {\it earlier} than the rest of the culture) and positive for the ranges~V to~VII (indicating that firing of these communities occurs on average {\it later} than the rest of the culture). The highest negative time delay is detected in range~III, such that the communities organized around its associated functional hubs can be described as local burst initiation cores~\cite{Eytan:2006hu, Eckmann:2010ii}.

\subsection*{Purely excitatory networks}

In this study we \JordiRev{did} not consider inhibitory interactions, neither in simulations nor in experiments (GABAergic transmission was blocked), \JordiRev{but we attempted uniquely the reconstruction of excitatory connectivity.}

We would like to point out that this is not a general limitation of TE, since the applicability of TE does not rely on assumptions as to the specific nature of a given causal relationship~-- for instance about whether a synapse is excitatory or inhibitory. In this sense, TE can be seen as a measure for the \emph{absolute} strength of a causal interaction, and is able in principle to capture the effects on dynamics of both inhibitory and excitatory connections. \DemianRev{Note indeed that previous studies \cite{Garofalo:2009jm, Ito:2011wc} have used TE to infer as well the presence of inhibitory connections.} However, TE alone could not discriminate the sign of the interaction \DemianRev{and additional \textit{post-hoc} considerations had to be made in order to separate the retrieved connections into separate excitatory and inhibitory subgroups (e.g., in \cite{Ito:2011wc}, based on the supposedly known existence of a difference in relative strength between excitatory and inhibitory conductances).}

\subsubsection*{\DemianRev{A simpler complex problem}}

\DemianRev{By focusing on excitatory connectivity only, our intention was} to simplify the full problem of network reconstruction, aiming as a first step to uncover systematically the strongest excitatory links in the network. \DemianRev{Such simplification allows indeed to remove potential causes of error, like, e.g.} disynaptic inhibition \DemianRev{or synchronous excitatory and inhibitory inputs to a same cell, that might result in failed inference of both the excitatory and the inhibitory connections. Nonetheless, this simpler problem remains very difficult because we attempt to reconstruct not only few connections but an entire adjacency matrix for the excitatory subnetwork.}

\subsubsection*{\DemianRev{Excitation shapes spontaneous activity}}

% Furthermore, we must say, first, that
\DemianRev{The inference of excitatory connectivity is by itself a very relevant issue. Excitatory recurrent connections are indeed a strong ---if not the main--- determinant of spontaneous activity.
% On one side, indeed, t
They act as an ``amplifier'', propagating to the network locally generated firing, and guaranteeing thus that the level of spontaneous activity of the culture remains elevated.}
% On the other side
\OlavRev{Furthermore,} \DemianRev{modeling studies (see e.g. \cite{Levina:2007ev, Levina:2009dn}) suggest that excitatory connections only are sufficient to obtain network bursts with experimentally observed statistics \cite{Beggs:2003uv, Mazzoni:2007jq}.}

\subsubsection*{\DemianRev{Sharper signals}}

\DemianRev{Second, when moving to experimental data,} \JordiRev{neurons fire more strongly when inhibition is blocked, which makes the identification of firing occurrences more accurate. The amplitude of the fluorescence signal increases by a factor two or more when inhibition is blocked (see e.g.~\cite{Jacobi:2009gn}). In Fig.~S8B we show an example of the fluorescence signal for the same neuron, before (blue) and after (black) blocking inhibitory synaptic transmission.} \DemianRev{Therefore the reconstruction problem of purely excitatory networks becomes simpler, not really from the algorithmic side (since the algorithm is potentially ready to cope with inhibition), but rather from the experimental side, because of an improved signal-to-noise ratio. Thus, when we generate synthetic data for algorithmic benchmarking, we aim at reproducing these same optimal experimental conditions.}

\JordiRev{We also note that the distribution of fluorescence values would be different for recordings with and without inhibition.} \DemianRev{Nevertheless, our method might still be applied, without need of qualitative changes, and state selection might still be performed. The optimal range for conditioning would be different and should again be estimated through a model-based benchmarking, but this would lead only to quantitative adjustments.}

\subsubsection*{\DemianRev{A two-steps strategy?}}

% If excitatory connectivity is crucial, it would be advisable however to reconstruct also inhibitory connectivity.
\DemianRev{A possible} strategy \OlavRev{to extend our method to the reconstruction of inhibitory connections} could be to follow a two-steps experimental approach: first, reconstruct only excitatory connectivity, based on recordings in which inhibitory transmission is blocked; and second, reconstruct only inhibitory connectivity, based on recordings in which, after the wash-out of the culture, excitatory transmission is blocked. In such an experiment, when recurrent excitation is blocked, the spontaneous level of firing activity should be restored by chemical non-synaptic activation. \OlavRev{We note} \DemianRev{however, that although complete blockade of excitation combined with drug-induced network activation is a standard protocol in slice studies (cfr. e.g.~\cite{Bartos:2007kl}), such an approach has never been attempted in cultures of dissociated neurons.}
% and we cannot commit on its feasibility.

\DemianRev{A compromise might be, therefore, to systematically compare the reconstructed connectivity before and after the wash-out of the inhibition blocker} \OlavRev{thereby} \DemianRev{collecting indirect evidence about the existence of inhibitory connection, through an analysis of the modulatory action they exert on the causal strengths of previously inferred excitatory connections.} \JordiRev{Note that, at least according to \cite{Ganguly:2001tz}, GABAergic synapses would be able to activate voltage dependent $\text{Ca}^{2+}$ channels and elevate the intracellular calcium concentration only before GABA switch (occurring around DIV 7, see later discussion about culture age). Hence, only before such early developmental event one could observe a modification of the fluorescence signal due to GABA action, making a two-steps approach necessary (first excitation, through direct analysis, and then, inhibition, indirectly through connectivity comparison).}

\JordiRev{Note also that the overall strength of inhibition in the network could be revealed by studying the decay in bursting activity as a result of gradually weaker excitation. Such an approach is possible by targeting the AMPA-glutamate receptors in excitatory neurons with increasing concentrations of antagonists such as CNQX (cfr.~\cite{Soriano:2008p747,Jacobi:2009gn}), and in the presence or absence of inhibitory synapses.} 

%% The first sentences are not totally true, so I prefer to drop them.

%\JordiRev{As stressed before, blocking of inhibition leads to a dynamical regime in which inter-burst activity is enhanced. %Reintroducing inhibition would therefore reduce the fraction of data that carries direct information about network topology. The %accuracy of the reconstruction method might therefore be affected.} 

\subsection*{\DemianRev{Experimental paradigm}}

\subsubsection*{\DemianRev{Connectivity in cultures vs. connectivity \textit{in vivo} }}

We have here used our current TE algorithm to the inference of connectivity from calcium fluorescence recordings of \emph{in vitro} cultured networks of dissociated cortical neurons. \DemianRev{However, such studies of cultures do not yield direct information about the connectivity of cortical tissues \textit{in vivo}.} \JordiRev{Several factors contribute to shape the neuronal circuits in an intact living brain, including the adequate orientation of neurons, dendrites, and axons, the biochemical guidance of processes towards their targets, and the refinement of circuitry through activity. Ultimately, development leads to a complex cortical structure organized both in layers and in columns, and with many particular topological features such as clusters or hierarchical organization \cite{Gong:2009fo}.}

\JordiRev{All these structures ---which are preserved in slices \cite{Song:2005p45, Lefort:2009bx, Stepanyants:2009bu, Perin:2011kp}--- are completely lost during the dissociation of the embryonic brain that precedes the preparation of \textit{in vitro} cultures. Neuronal cultures form} \DemianRev{new connections ``from scratch''}, \JordiRev{with a combination of short and long length connections, leading to circuits that have orders of magnitude less neurons and synapses.} 

\DemianRev{Nevertheless, understanding how neurons wire together spontaneously in a controlled medium is also of utmost importance, because it allows separating endogenous and exogenous driving components of neuronal wiring.} \JordiRev{Furthermore, cultures and slices share similar spontaneous bursting dynamics.} 
\DemianRev{If this observation alone should not be used to support the existence of shared topological features (cfr. Fig.~\ref{fig:CCnets_all_look_alike}),} \JordiRev{it is true that the self-organization principles underlying the development of networks up to a bursting dynamical state may be common in all living neuronal networks.}

\subsubsection*{\DemianRev{Calcium imaging vs. multielectrode arrays}}

Our algorithm has been optimized for the application to real calcium imaging data, by determining an optimal conditioning range based only on qualitative features of the distribution of the average fluorescence in the network (very similar for synthetic and real data). \DemianRev{Other studies have however used electrophysiological recordings from cultures grown on multielectrode arrays (MEAs) \cite{Eckmann:2007ft, Marom:2002va,Beggs:2003uv,Wagenaar:2006fo,Erickson20081} as a basis for their topology reconstruction strategies (see e.g. \cite{Garofalo:2009jm, Ito:2011wc}).} 

\JordiRev{MEAs provide excellent temporal resolution and, to a certain extent, also the possibility to resolve individual spikes. However, MEAs have a limited number of electrodes and often neurons are not precisely positioned on an electrode but at its vicinity, which requires complex processing of the data to identify the source of a given spike. Additionally, the electrodes have to be in contact or embedded into the neuronal tissue, limiting its use to mostly cultures and brain slices.}

\JordiRev{Calcium imaging, in contrast, offers important advantages. First, the technique provides access to the activity of thousands of neurons in large fields of view for several hours \cite{Soriano:2008p747,Feinerman:2008vc}, and with a time resolution} \DemianRev{proven to be sufficient for reconstruction in the present study. Second, calcium imaging with superior temporal and spatial resolution~\cite{Takano:2012jw,Grienberger2012} is a technique that combines perfectly with new breakthroughs in experimental neuroscience, particularly optogenetics \cite{Yizhar:2011jv,Knopfel2010} and genetically encoded calcium indicators \cite{Hires2008,Knopfel2010}, technologies that are under expansion both \emph{in~vitro} \cite{Li06122005} and \emph{in~vivo} \cite{Buchen2010}.} \DemianRev{Our reconstruction method can promptly be readapted for the analysis of other calcium imaging datasets.}\JordiRev{ The analysis of \emph{in~vitro} cultures of dissociated neurons is just a first proof-of-concept of the applicability of generalized TE to real data.} 

%Note that, at least for what concerns reconstructions based on surrogate data, our method based on poor time-resolution calcium signals (rather than on exact spike times) out-performed \cite{Garofalo:2009jm} and reached comparable performance levels to \cite{Ito:2011wc}.}

\DemianRev{As a matter of fact, extended TE might even be applied to multielectrode array data with very few modifications, along lines analogous to \cite{Ito:2011wc}. The advantages of an increased time-resolution might then be combined with the efficacy of the state-selection through conditioning concept.}

\subsubsection*{\DemianRev{Age of neuronal cultures}}

\DemianRev{In our study} \OlavRev{we recorded from} \DemianRev{\textit{early mature} (DIV 9-12) instead of fully mature cultures. Young but sufficiently mature cultures have rich bursting events while preserving some isolated activity . On the contrary, fully mature cultures show strong synchronized bursting dynamics (epileptic-like), with very little isolated neuronal activity \cite{Kamioka1996}. In this sense, therefore, young cultures emerge as a model system which better matches the features of non pathologic cortical tissue activity. At the same time, conditions are ideal for an analysis focusing on inter-burst periods, as our one (different, in this sense, from an alternative approach focusing on burst initiation as \cite{Pajevic:2009p252}.)} 

\JordiRev{Several authors have studied the process of maturation of neuronal cultures, and characterized their spontaneous activity along days or weeks \cite{vanPelt:2004kc,Tetzlaff:2010he}. Some studies have identified a stage of full maturation and stable bursting dynamics at the third week, a stage that can last for months (e.g. \cite{Kamioka1996}).} \JordiRev{However, depending on neuronal density \cite{Wagenaar:2006fo,Cohen:2008cv, Soriano:2008p747}, glial density \cite{Feldt2010}, and the gestation time of the embryo at the moment of dissection \cite{Soriano:2008p747}, spontaneous activity with rich episodes of population bursts} \DemianRev{can emerge as early as DIV 3-4 \cite{Cohen:2008cv,Soriano:2008p747}).}

\DemianRev{On the other hand, GABA switch (the developmental event after which the action of GABAergic synapses become inhibitory as in fully developed networks and stops being excitatory \cite{Cherubini:1991ws})} \JordiRev{in cultures similar to ours occurs at around day 7 \cite{Cohen:2008cv, Soriano:2008p747}. Therefore, it has already taken place in early mature cultures at DIV~9 (such as the one analyzed in Fig.~S7). This is confirmed by the fact that the blockade of inhibition by bicuculline leads to an increase of the amplitude of the fluorescent signal by a factor 2-3 (see Figure~S8B), as expected after GABA switch has occurred.}

\subsection*{\DemianRev{Non-local and moderately clustered connectivity in cultures}}

\subsubsection*{\DemianRev{Evidence for long-range connections}}

\DemianRev{Our TE-based algorithm applied to the reconstruction of the topology of neuronal cultures in vitro have inferred the existence of direct connections between neurons separated by a considerable distance. Indeed,} the reconstructed distribution of connection distances peaks at a remarkably high value and is markedly high-skewed (Fig.~\ref{fig:real_data}B) 

Experimental studies showed that cortical slices have a maximum probability of connection at much shorter distances~\cite{Holmgren:2003p651}. We note, however, that the density of cells in our culture is considerably more diluted than in normal cortical tissues. Furthermore, cortical developmental processes strongly dictate \textit{in vivo} (and slice) connectivity \cite{Price:2006hw}, while a larger freedom to establish connections exist potentially in cultures \textit{in vitro}. \JordiRev{The growth process and the final length of the processes depend not only on the density of the culture and the population of glia, but also on the chemical properties of the substrate where neurons and process grow \cite{Grumbacher-Reinert01091989}.} The lack of restrictions during the development of neuronal processes (axons and dendrites) leads to longer axonal lengths in the culture on average, explaining the high connection distance obtained in the reconstruction. 

\JordiRev{Such long axons in cultures have been reported in other studies~\cite{Erickson20081,Kriegstein:1983uj,Feinerman:2005ka}, providing broad distributions of lengths with an average value of $1.1~\mbox{mm}$. Using Green Fluorescence Protein transfection\cite{Zeitelhofer2007}, we have directly visualized axonal processes as long as $1.8~\mbox{mm}$ in cultures of comparable age and density to the ones that we used for the fluorescence imaging recordings (Figure~S8C). Such long axons would be required to mediate the long-range causal interactions captured by the TE analysis. As a matter of fact, as revealed by the randomization of the reconstructed connectivity, long average connection lengths might simply be expected to exist as a result of the bi-dimensional distribution of neurons on a substrate in combination with long individual axonal lengths (i.e. spanning the entire field of view).}

For a younger culture at 9~DIV (Fig.~S7), on the other hand, we retrieved a probability of connection at short distances higher than expected from randomized networks. This over-connectivity at short distances can be ascribed to the fact that young cultures have less developed axons and therefore a connectivity favored towards closer neurons. However, more exhaustive studies ---going beyond the scope of the present work--- would be needed to assess the full dependence of reconstructed network topology on culture age, neuronal density and spatial distribution, as well as on connection length.

\subsubsection*{\DemianRev{Evidence for moderate clustering}}

Our TE-based analysis suggest an enhanced tendency to clustering of connections in the analyzed neuronal cultures. \DemianRev{Indeed, although the average clustering coefficient is moderated, it is significantly higher than what might be expected based on randomized networks.}

Neurons in culture aggregate during growth, giving rise to fluctuations in neuronal density. Since denser areas in the culture have been reported to have a higher connectivity \cite{Soriano:2008p747}, some level of clustering is naturally expected in the real network, as detected by the reconstruction. We could not however identify in our reconstructed topology a statistically significant correlation between the local density of neurons ---quantified by the number of cells within a radius of $50 \mu\mathrm{m}$ centered on each given neuron--- and the local degree of clustering ($r < 0.01$ for both considered cultures). 

This finding suggests the existence of a mixture of local and non-local clustering in the culture and indicates that network clustering in a sparse culture is not a mere byproduct of the inhomogeneous density of cells, but might reflect activity-dependent mechanisms for synaptic wiring.

\subsubsection*{\DemianRev{A bias of cross-correlation methods?}}

XC-based reconstructions inferred a much more local average distance of connection, significantly smaller not only than the inference based on TE, but also than what expected from randomized networks. At the same time, XC-based methods inferred an average clustering coefficient almost twice as large as TE approaches.

As a matter of fact, the analyses of Fig.~\ref{fig:one_reconstruction_illustrated-CC}~and~\ref{fig:one_reconstruction_illustrated-Lambda} suggest that the connectivity of real cultures inferred by XC is prone to include artifactual features, such as an exceedingly local connectivity, paired to an overestimated level of clustering \DemianRev{with respect to ground-truth (unknown in cultures).}

\DemianRev{Interestingly (and possibly indicative of biased estimations), the discrepancies between TE-based and XC-based inferences of mean connection distance and clustering level are paralleling both the systematic deviations between XC-bases and TE-based reconstructions identified in applications to surrogate data.}

\subsubsection*{\OlavRev{No} \DemianRev{evidence for scale free connectivity}}

Despite the observation that the distribution of degrees inferred by TE is characteristically broadened and has a conspicuous right tail, we have found no evidence of a scale free degree distribution \DemianRev{(i.e. following a power-law \cite{Barabasi:1999p20}).} \JordiRev{A similar conclusion was also reached for cultured neuronal networks in Ref.~\cite{Srinivas:2007jg}.}

On the contrary, studies like~\cite{Pajevic:2009p252} hinted at a large overlap in cultures between the structural connectivity and the retrieved scale free functional connectivity, \JordiRev{meeting other studies that identified power-law degree distributions in the functional connectivity of cultures \cite{Eytan:2006p66}} \DemianRev{or of slices~\cite{Bonifazi:2009in}.}

\JordiRev{Whether connectivity in neural networks, \textit{in vitro} or \textit{in vivo} can be characterized as scale free or not is a highly debated issue (see e.g.~\cite{Kaiser:2011gf}),} \DemianRev{and, especially for self-organized networks of dissociated neurons, it is likely to depend on details of how the culture is grown.} \DemianRev{The connectivity retrieved from our calcium recordings might possibly be better described as ``small-world''~\cite{Watts:1998p78}, due to the existence of hub nodes with very high out- and in-degree. However, due to the large uncertainty in the reconstruction, we did not attempt any systematic assessment of the average path length between nodes (since this is a quantity very sensitive to error), and prefer to describe the reconstructed degree distribution just as ``right-skewed''.}

\subsubsection*{\DemianRev{How reliable are absolute values of reconstructed properties?}}

\DemianRev{Through our reconstructions of culture connectivity, we have provided actual absolute values for properties such as average degree, average clustering or average connection length.} \OlavRev{We note that such estimates} \DemianRev{are affected by a large uncertainty that goes beyond the variability described by their reconstructed distributions. Indeed, these estimates come from a reconstruction based on a specific choice of the number of links to include. This choice, as highlighted by Figure~S1 for synthetic data, corresponds to selecting directly a specific value of the average culture in- and out-degree. Therefore, as previously discussed, the included link number should be determined by our expectations on the average degree of the network to reconstruct, based on independent experimental evidence or on extrinsic guiding hypotheses.}

\DemianRev{The inference of average clustering or of average distance is robust even against relatively large mistakes in the initial guess for the average degree (see e.g. Figure~S1, for clustering estimation). In the case of the DIV 12 network reconstructed in Fig.~\ref{fig:real_data}, doubling the threshold of included links from 5\% to 10\% (and therefore adopting a twice as large guess for the average culture degree) changes the inference for the average clustering coefficient from $0.090 \pm 0.010$ to $0.157 \pm 0.011$ and the average connection distance from $852 \pm 463 \mu$m to $855 \pm 464 \mu$m. Nevertheless, the precise values obtained do depend on the} \OlavRev{number of reconstructed links.}

\DemianRev{Due to the lack of information, a better strategy might be, rather than focusing on absolute estimated values, to focus on comparisons with fully and partially randomized networks with analogous average degree or degree distribution, respectively. Such comparisons indeed can convey qualitative information about the occurrence of non-trivial deviations from chance expectations, which are likely to be more reliable than quantitative assessments.}

\subsection*{Conclusions and perspectives}

In summary, we have developed a new generalization of Transfer Entropy for inferring connectivity in neuronal networks based on fluorescence calcium imaging data. Our new formalism goes beyond previous approaches by introducing two key ingredients, namely the inclusion of \emph{same bin} interactions and the separation of dynamical states through \emph{conditioning} of the fluorescence signal. We have thoroughly tested our formalism in a number of simulated neuronal architectures, and later applied it to extract topological features of real, cultured cortical neurons.

We expect that, in the future, algorithmic approaches to network reconstruction, and in particular our own method, will play a pivotal role in unravelling not only topological features of neuronal circuits, but also in providing a better understanding of the circuitry underlying neuronal function. These theoretical and numerical tools may well work side by side with new state-of-the-art techniques (such as optogenetics or high-speed two-photon imaging~\cite{Grinvald:2004p5542,Wang:2005p5539,Holekamp:2008p5557,Chemla:2010p5541,Bernstein:2011ec}) that will enable direct large-scale reconstructions of living neuronal networks. Our Transfer Entropy formalism is highly versatile and could be applied to the analysis of \emph{in vivo} voltage-sensitive dye recordings with virtually no modifications.

On a shorter time-scale, \DemianRev{it would be important to extend our} analysis to the reconstruction of both excitatory and inhibitory connectivity in \emph{in vitro} cultures, which is technically feasible, and to compare diverse network characteristics, such as neuronal density or aggregation. \DemianRev{Our algorithm could be used} to systematically \DemianRev{reconstruct the connectivity of cultures at} different development stages in the quest for understanding the switch from local to global neuronal dynamics. Another crucial \DemianRev{open issue is to design suitable} experimental protocols \DemianRev{allowing to confirm the existence of at least some of the inferred synaptic links, in order to} validate statistically the reconstructed connectivity. For instance, \DemianRev{the actual presence of directed links to which our algorithm assigns the largest TE scores might be systematically probed through targeted} paired electrophysiological stimulation and recording. \DemianRev{Furthermore,} GFP transfection or inmunostaining \DemianRev{might be used} to obtain actual, precise anatomical data on network architecture to be compared with the reconstructed one. Finally, \DemianRev{it might be interesting} to reconstruct connectivity of cultured networks before and after physical disconnection of different areas of the culture (e.g. by mechanical etching of the substrate or by chemical silencing). These manipulations would provide a scenario to verify whether TE-based reconstructions correctly capture the absence of direct connections between areas of the neuronal network which are known to be artificially segregated.

% You may title this section "Methods" or "Models".
% "Models" is not a valid title for PLoS ONE authors. However, PLoS ONE authors may use "Analysis"
\section*{Materials and Methods}

\subsection*{Network construction and topologies}

We generated synthetic networks with~$N=100$ neurons, distributed randomly over a squared area of~0.5mm lateral size. We chose~$p=0.12$ as the connection probability between neurons \cite{Wen28072009}, leading to sparse connectivities similar to those observed in local cortical circuits~\cite{Song:2005p45}. We used non-periodic boundary conditions to reproduce eventual ``edge'' effects that arise from the anisotropic cell density at the boundaries of the culture.

We considered two general types of networks: (i) a {\it locally-clustered ensemble}, where the probability of connection depended on the spatial distance between two neurons; and (ii) a {\it non-locally clustered ensemble}, with the connections engineered to display a certain degree of clustering.

For the case of a non-local clustering ensemble, we first created a sparse connectivity
matrix, randomly generating links with a homogeneous probability of connection across
pairs of neurons. We next selected a random pairs of links and ``crossed'' them (links $A
\rightarrow B$ and $C \rightarrow D$ became $A \rightarrow D$ and $C \rightarrow B$). We
accepted only those changes that updated the clustering index in the direction of a
desired target value, thereby maintaining the number of incoming as well as outgoing
connections of each neuron. The crossing process was iterated until a clustering index
higher or equal to the target value was reached. The overall procedure led to a full
clustering index of the reference random network of $0.120 \pm 0.004$ (mean and standard
deviation, respectively, across 6 networks). After the rewiring iterations, we then
achieved standard deviations from the desired target clustering value smaller than~0.1\%
for all higher clustering indices.

We measured the full clustering index of our directed networks according to a common definition introduced by~\cite{Fagiolo:2007p53}:
\begin{equation}
    \text{CC} = \langle \frac{(A+A^T)_{ii}^3}{2 d_i^{\text{tot}} (d_i^{\text{tot}}-1) - 4 d_i^{\text{bidir}} }
    \rangle_i.
\end{equation}
The binary adjacency matrix is denoted by~$A$, with $A_{ji}=1$ for a link~$j \to i$, and zero otherwise. The adjacent matrix provides a complete description of the network topological properties. For instance, the in-degree of a node~$i$ can be computed as $d_i^{\text{in}} = \sum_j A_{ji}$, and the out-degree as $d_i^{\text{out}} = \sum_j A_{ij}$. The total number of links of a node is given by the sum of its in-degree and its out-degree ($d_i^{\text{tot}} = d_i^{\text{in}} + d_i^{\text{out}}$). The number of bidirectional links of a given node~$i$ (i.e. links between~$i$ and~$j$ so that~$i$ and~$j$ are reciprocally connected by directed connections) is given by~$d_i^{\text{bidir}} = (A)_{ii}^2$.

The adjacency matrix did not contain diagonal entries. Such entries would correspond to``autaptic'' links that connect a neuron with itself. Note that our \DemianRev{directed functional} connectivity analysis is based on bivariate time series, and therefore it would be structurally unfit to detect this type of links.

For the case of the local clustering ensemble, two neurons separated a Euclidean distance~$r$ were randomly connected with a distance dependent probability described by a Gaussian distribution, of the form~$p_0(r) = \exp(-(r/\lambda)^2)$, with~$\lambda$ a characteristic length scale. To guarantee that a constant average number of links~$C$ was present in the network, this Gaussian distribution was rescaled by a constant pre-factor, obtained as follows. We first generated a network based on the unscaled kernel $p_0(r)$ and computed the resulting number of links~$C'$. With this value we then generated a final network based on the rescaled kernel~$p(r) = C/C' \exp(-(r/\lambda^2))$.

\subsection*{Simulation of the dynamics of cultured networks}

The dynamics of the generated neuronal networks was studied using the NEST simulator~\cite{Gewaltig:NEST,Eppler:2008p3234}. We modeled the neurons as leaky integrate-and-fire neurons, with the membrane potential~$V_i(t)$ of a neuron~$i$ described by~\cite{Tsodyks:1997p5532,Tsodyks:2000p3237}:
\begin{equation}
    \tau_m \frac{d V_i(t)}{d t} = -V_i + \frac{I_{\text{syn}}(t)}{g_l},
\end{equation}
where~$g_l=50\text{pS}$ is the leak conductance and $\tau_m=20\text{ms}$ is the membrane time-constant. The term $I_{\text{syn}}$ account\DemianRev{s} for a time-dependent input current that arises from recurrent synaptic connections. In the absence of synaptic inputs, the membrane potential relaxes exponentially to a resting level set arbitrarily to zero. Stimulation in the form of inputs from other neurons increase the membrane potential, and above the threshold~$V_{\text{thr}}=20\text{mV}$ an action potential is elicited (\emph{neuronal firing}). The membrane voltage is then reset to zero for a refractory period of~$t_{\text{ref}}=2\text{ms}$.

The generated action potential excites post-synaptic target neurons. The total synaptic currents are then described by
\begin{equation}
    \frac{d I_{\text{syn}}(t)}{d t} = -\frac{I_{\text{syn}}}{\tau_s} + \alpha_{\text{int}} \displaystyle \sum_{j=1}^N \sum_k A_{ji} \, E_{ji}(t) \, \delta(t-t_j^k-t_d) + \alpha_{\text{ext}} \displaystyle \sum_l \delta(t-t_{\text{ext},i}^l-t_d), \label{eq:LIF}
\end{equation}
where~$A$ is the adjacency matrix, and $\tau_s=2\text{ms}$ is a synaptic time constant. The resulting excitatory post-synaptic potentials~(EPSPs) have a standard difference-of-exponentials time-course \cite{Dayan:2005uy}.

Neurons in culture show a rich spontaneous activity that originates from both
fluctuations in the membrane potential and small currents in the pre-synaptic terminals
(\emph{minis}). The latter is the most important source of noise and plays a pivotal role
in the generation and maintenance of spontaneous activity \cite{Cohen:2011bi}. To
introduce the spontaneous firing of neurons in Eq.~(\ref{eq:LIF}), each neuron $i$ was
driven, through a static coupling conductance with strength
$\alpha_{\text{ext}}=4.0\text{pA}$, by independent Poisson spike trains (with a
stationary firing rate of $\nu_{ext}=1.6\text{Hz}$, spikes fired at stochastic times
$\{t_{\text{ext},i}^l\}$).

Neurons were connected via synapses with short-term depression, due to the finite
amount of synaptic resources~\cite{Tsodyks:2000p3237}. We considered only purely
excitatory networks to mimic the experimental conditions in which inhibitory transmission
is fully blocked. Concerning the recurrent input to neuron~$i$, the set~$\{ t_j^k \}$
represents times of spikes emitted by a presynaptic neuron~$j$, $t_d$~is a conduction
delay of~$t_d=2\text{ms}$, while~$\alpha_{\text{int}}$ sets a homogeneous scale for the
synaptic weights of recurrent connections, whose time-dependent
strength~$\alpha_{\text{int}} E_{ji}(t)$ depends on network firing history through the
equations
\begin{eqnarray}
    \frac{d E_{ji}(t)}{d t} &=& -\frac{E_{ji}}{\tau_{\text{inact}}} + U \, R_{ji} \displaystyle \sum_k \delta(t-t_j^k), \\
    \frac{d R_{ji}(t)}{d t} &=& -\frac{1}{\tau_{\text{rec}}} \left( 1-R_{ji}-E_{ji} \right).
\end{eqnarray}
In these equations, $E_{ji}(t)$ represents the fraction of neurotransmitters in the ``effective state'', $R_{ji}(t)$ in the ``recovered state'' and $I_{ji}(t) = 1-R_{ji}-E_{ji}$ in the ``inactive state''~\cite{Tsodyks:1997p5532,Tsodyks:2000p3237}. Once a pre-synaptic action potential is elicited, a fraction~$U=0.3$ of the neurotransmitters in the recovered state enters the effective state, which is proportional to the synaptic current. This fraction decays exponentially towards the inactive state with a time scale~$\tau_{\text{inact}}=3\text{ms}$, from which it recovers with a time scale~$\tau_{\text{rec}}=500\text{ms}$. Hence, repeated firing of the presynaptic cell in an interval shorter than~$\tau_{\text{rec}}$ gradually reduces the amplitude of the evoked EPSPs as the synapse is experiencing fatigue effects (depression).

Random networks of integrate-and-fire neurons coupled by depressing synapses are
well-known to naturally generate synchronous events~\cite{Tsodyks:2000p3237}, comparable
to the all-or-none behavior that is observed in cultured
neurons~\cite{Eytan:2006p66,Eckmann:2008p77}. To obtain in our model a realistic bursting
rate~\cite{Eckmann:2008p77}, the synaptic weight of internal connections was set to
result into a network bursting of $0.10 \pm 0.01\text{Hz}$ for all the network
realizations that we studied, and in particular for any considered (local or non-local)
clustering level. Therefore, after having generated each network topology, we assigned
the arbitrary initial value of $\alpha_{\text{int}} = 5.0\text{pA}$ to internal synaptic
weights and simulated 200~seconds of network dynamics, evaluating the resulting average
bursting rate. If it was larger (smaller) than the target bursting rate, then the
synaptic weight $\alpha_{\text{int}}$ was reduced (increased) by~10\%. We then
iteratively adjusted~$\alpha_{\text{int}}$ by (linearly) extrapolating the last two
simulation results towards the target bursting rate, until the result was closer
than~0.01Hz to the target value. The resulting used values of~$\alpha_{\text{int}}$ are
provided in Table~\ref{tab:synaptic_weights_used_for_simulation}. 

Note that we defined a network burst to occur when more than~40\% of the neurons in the network were active
within a time window of~50ms. \DemianRev{Such a criterion does not play any role in the reconstruction algorithm itself, where state selection is achieved through conditioning, but is only used for the automated generation of random networks with a prescribed bursting rate. Typically, for a fully developed burst, more than 90\% of the neurons fire within a 50 ms bin, while, during inter-burst intervals, less than 10\% do. Due to the clear separation between these two regimes, our burst detection procedure does not depend significantly on the precise choice of threshold within a broad interval.}

\subsection*{Model of calcium fluorescence signals}

To reproduce the fluorescence signal measured experimentally, we treated the simulated spiking dynamics to generate surrogate calcium fluorescence signals. We used a common model introduced in~\cite{Vogelstein:2009p3029} that gives rise to an initial fast increase of fluorescence after activation, followed by a slow decay~($\tau_{\text{Ca}}=1\text{s}$). Such a model describes the intra-cellular concentration of calcium that is bound to the fluorescent probe. The concentration changes rapidly by a step amount of~$A_{\text{Ca}} = 50 \mu\text{M}$ for each action potential that the cell is eliciting in a time step~$t$, of the form
\begin{equation}
    [\text{Ca}^{2+}]_t - [\text{Ca}^{2+}]_{t-1} = - \frac{\Delta t}{\tau_{\text{Ca}}} [\text{Ca}^{2+}]_{t-1} + A_{\text{Ca}} \, n_t,
\end{equation}
where~$n_t$ is the total number of action potentials.

The net fluorescence level~$F$ associated to the activity of a neuron~$i$ is finally obtained by further feeding the Calcium concentration into a saturating static non-linearity, and by adding a Gaussian distributed noise~$\eta_t$ with zero mean:
\begin{equation}
    F_{i,t} = \frac{[\text{Ca}^{2+}]_t}{[\text{Ca}^{2+}]_t + K_d} + \eta_t.
\end{equation}
For the simulations, we used a saturation concentration of~\JordiRev{$K_d = 300 \mu\text{M}$} and noise with a standard deviation of~0.03.

\subsection*{Modeling of light scattering}

We considered the light scattered in a simulated region of interest~(ROI) from surrounding cells. Denoting as~$d_{ij}$ the distance between two neurons~$i$ and~$j$ and by~$\lambda_{sc}=0.15~\text{mm}$ the scattering length scale (determined by the typical light deflection in the medium and the optical apparatus), the resulting fluorescence amplitude of a given neuron~$F^{\tt{sc}}_{i,t}$ is given by
\begin{equation}
    F^{\tt{sc}}_{i,t} = F_{i,t} + A_{sc} \displaystyle \sum_{j=1, j \neq i}^N F_{j,t} \exp \left\{ - \left(d_{ij}/\lambda_{sc}\right)^2
    \right\}.
\end{equation}

A sketch illustrating the radius of influence of the light scattering phenomenon is given
in Fig.~S9. The scaling factor~$A_{sc}$ sets the overall strength of the simulated
scattering artifact. Note that light scattered, according to the equation shown above,
could be completely corrected using a standard deconvolution algorithm, at least for very
large fields of view and a scattering length known with sufficient accuracy. In a real
setup however, the relatively small fields of view (on the order of 3 mm$^2$), the
inaccuracies in inferring the scattering radius~$\lambda_{sc}$, as well as the
inhomogeneities in the medium and on the optical system, make perfect deconvolution not
possible. Therefore, artifacts due to light scattering cannot be completely
eliminated~\cite{Ishimaru:1999wx,Minet:2008us}. The scaling factor~$A_{sc}$, that we
arbitrarily assumed to be small and with value $A_{sc}=0.15$, can be seen as a measure of
this residual artifact component.

\subsection*{Generalized Transfer Entropy}

In its original formulation~\cite{Schreiber:2000p79}, for two discrete Markov processes~$X$ and~$Y$ (here shown for equal Markov order~$k$), the Transfer Entropy (TE) from~$Y$ to~$X$ was defined as:
\begin{equation}
    \mbox{TE}_{Y \rightarrow X} = \displaystyle \sum P(x_{n+1}, x^{(k)}_n, y^{(k)}_n) \, \log \frac{P(x_{n+1} | x^{(k)}_n, y^{(k)}_n)}{P(x_{n+1} | x^{(k)}_n)}, \label{eq:TE-def}
\end{equation}
where~$n$ is a discrete time index and~$x^{(k)}_n$ is a vector of length~$k$ whose entries are the samples of $X$ at the time steps $n$,~$n-1$, ...,~$n-k$. The sum goes over all possible values of~$x_{n+1}$, $x^{(k)}_n$ and~$y^{(k)}_n$.

TE can be seen as the distance in probability space (known as the Kullback-Leibler divergence~\cite{MacKay:2003wc}) between the ``single node'' transition matrix $P(x_{n+1} | x^{(k)}_n)$ and the ``two nodes'' transition matrix $P(x_{n+1} | x^{(k)}_n, y^{(k)}_n)$. As expected from a distance measure, TE is zero if and only if the two transition matrices are identical, i.e. if transitions of~$X$ do not depend statistically on past values of~$Y$, and is greater than zero otherwise, signaling dependence of the transition dynamics of~$X$ on~$Y$.

We use TE to evaluate the \DemianRev{directed functional} connectivity between different network nodes. In a pre-processing step, we apply a basic discrete differentiation operator to calcium fluorescence time series~$F^{\tt{sc}}_{x,t}$, as a rather crude way to isolate potential spike events. Thus, given a network node~$x$, we define $x_n = F^{\tt{sc}}_{x,n+1}-F^{\tt{sc}}_{x,n}$. This pre-processing step also improves the signal-to-noise ratio, thus allowing for a better sampling of probability distributions with a limited number of data points. To adapt TE to our particular problem we need to take into account the general characteristics of the system. We therefore modified TE in two crucial aspects:
\begin{enumerate}
    \item We take into account that the synaptic time constants of the neuronal network ($\sim$1 ms) are much shorter than the acquisition times of the recording ($\sim$10 ms). We therefore need to account for ``same bin'' causal interactions between nodes, i.e. between events that fall in the same time-bin. Slower interactions with longer lags are still taken into account by evaluating TE for a Markov order larger than one (in time-bins units).
    \item We consider the possibility that the network dynamics switches between multiple dynamical states, i.e.~between bursting and inter-bursting regimes. These regimes are characterized by different mean rates of activity and, potentially, by different transition matrices. Hence, we have to restrict the evaluation of TE to time ranges in which the network is consistently in a single dynamical state. The separation of dynamical states can be achieved by introducing a variable~$g_t$ for the average signal of the whole network,
    \begin{equation}
        g_t = \frac{1}{N} \displaystyle \sum_{i=1}^N x_i(t).
    \end{equation}
    We then include all data points at time instants in which this average fluorescence~$g_t$ is below a predefined threshold parameter~$\tilde{g}$, i.e. we consider only the time points that fulfill $ \{ t : g_t < \tilde{g} \} $. We only make an exception that corresponds to the simulations of figures~\ref{fig:state_dependency} and~S2, where we considered time points that fall within an interval bounded by a higher and a lower thresholds, i.e. $ \{ t : \tilde{g}_{\text{low}} < g_t < \tilde{g}_{\text{high}} \} $.
\end{enumerate}

Using these two novel aspects, we have extended the original description of Transfer
Entropy [Eq.~(\ref{eq:TE-def})] to the following form
\begin{equation}
    \mbox{TE}^*_{Y \rightarrow X}(\tilde{g}) = \displaystyle \sum P(x_{n+1}, x^{(k)}_n, y^{(k)}_{n+1} | g_{n+1}<\tilde{g}) \, \log \frac{P(x_{n+1} | x^{(k)}_n, y^{(k)}_{n+1}, g_{n+1}<\tilde{g})} {P(x_{n+1} | x^{(k)}_n, g_{n+1}<\tilde{g})}. \label{eq:ourTE-def}
\end{equation}

Probability distributions have to be evaluated as discrete histograms. Hence, the continuous range of fluorescence values (see e.g. the bottom panels of Fig.~\ref{fig:raw_data_vs_simulations}) is quantized into a finite number~$B$ of discrete levels. We typically used a small~$B = 3$, a value that we justify based on the observation that the resulting bin width~$b$ is close to twice the standard deviation of the signal. The presence of large fluctuations, most likely associated to spiking events, is then still captured by such a coarse, almost non-parametric description of fluorescence levels.

\subsection*{Network reconstruction}

Generalized TE values are obtained for every possible directed pair of network nodes, and using a fixed threshold level $\tilde{g}$. The set of TE scores are then ranked in ascending order and scaled to fall in the unit range. A threshold TE$_{thr}$ is then applied to the rescaled data, so that only those links with scores above~TE$_{\text{thr}}$ are retained in the reconstructed network.

A standard \textit{Receiver-Operator Characteristic} (ROC) analysis is used to assess the
quality of the reconstruction by evaluating the number of true positives (reconstructed
links that are present in the actual network) or false positives (not present), and for
different threshold values~TE$_{\text{thr}}$~\cite{Fawcett:2006um}. The highest threshold
value leads to zero reconstructed links and therefore zero true positives and false
positives. At the other extreme, the lowest threshold provides both~100\% of true
positives and false positives. Intermediate thresholds give rise to a smooth curve of
true/false positives as a function of the threshold. The performance of the
reconstruction is then measured as the degree of deviation of this curve from the
diagonal, and that corresponds to a random choice of connections between neurons.

To provide a simple method to compare different reconstructions, we arbitrarily use the quantity~TP$_{10\%}$, defined as the fraction of true positives for a ~10\% of false positives, as indicator for the quality of the reconstruction.

\OlavRev{An alternative to the ROC is the \textit{Positive Prediction Curve}~\cite{Garofalo:2009jm}, plotting the ``true-false ratio'' (TFR) against the number of reconstructed links, called ``true-false sum'' (TFS). The TFR represents the fraction of true positives relative to the false positives. Denoting by~\#TP the absolute number of true positives and by~\#FP the number of false positives, TFR is therefore defined in~\cite{Garofalo:2009jm} in the following way:
\begin{equation}
  \text{TFR} = \frac{\text{\#TP} - \text{\#FP}}{\text{\#TP} + \text{\#FP}}
\end{equation}
The case $\text{TFR} = 0$ corresponds to the case that, for any given reconstructed link, it is on average equally likely that it is in fact a true positive rather than a false positive.
}

\subsection*{Alternative reconstruction methods}

To gain further insight into the quality of our reconstruction method, we compare reconstructions based on TE with three other reconstruction strategies, namely cross-correlation, mutual information, and Granger causality.

Cross-correlation~(XC) reconstructions are based on standard Pearson cross-correlation. The score assigned to each potential link is given by the largest cross-correlogram peak for lags between~$0$ and $t_{max}= 60$ms, of the form
\begin{equation}
    \mbox{XC}_{Y \rightarrow X} = \max_{\Delta t = 0...t_{max}} \left\{ \mbox{corr} \left( x^{(S-\Delta t)}_{S}, y^{(S-\Delta t)}_{S-\Delta t} \right) | g_S<\tilde{g}
    \right\}.
    \label{eq:Xcorr-def}
\end{equation}

In a similar way, the scores for Mutual Information~(MI) reconstructions are evaluated as
\begin{equation}
    \mbox{MI}_{Y \rightarrow X} = \max_{\Delta t = 0...t_{max}} \left\{ \displaystyle \sum P(x_n, y_{n-\Delta t} | g_n<\tilde{g}) \, \log \frac{P(x_n, y_{n-\Delta t} | g_n<\tilde{g})} {P(x_n | g_n<\tilde{g}) \, P(y_{n-\Delta t} | g_n<\tilde{g})} \right\}. \label{eq:MI-def}
\end{equation}
Analogously to TE, the sum goes over all entries of the joint probability matrix.

For the reconstruction based on Granger causality (GC)~\cite{Granger:1969p4641} we first model the signal~$x_t$ by least-squares fitting of a univariate autoregressive model, obtaining the coefficients~$a_k^0$ and the residual~$\eta^0$,
\begin{equation}
    x_t = \displaystyle \sum_{l=1}^k a_l^0 \: x_{t-l} + \eta_t^0 \label{eq:GC-pre-def1}.
\end{equation}
In a second step, we fit a second bivariate autoregressive model that includes the potential source signal~$y_t$, and determine the residual~$\eta^1$,
\begin{equation}
    x_t = \sum_{l=1}^k a_l^1 \: x_{t-l} + \sum_{m=0}^{k-1} b_m^1 \: y_{t-m} + \eta_t^1. \label{eq:GC-pre-def2}
\end{equation}
Note that in the latter bivariate regression scheme we take into account ``same bin''
interactions as for Transfer Entropy (index of the second sum starts at $m=0$).
Given~$\Gamma^0$, the covariance matrix of the univariate fit in
Eq.~(\ref{eq:GC-pre-def1}), and~$\Gamma^1$, the covariance matrix of the bivariate fit in
Eq.~(\ref{eq:GC-pre-def2}), GC is then given by the logarithm of the ratio between their
traces:
\begin{equation}
    \text{GC}_{Y \rightarrow X} = \log \frac{(\Gamma^0)_{0,0} + (\Gamma^0)_{1,1}}{(\Gamma^1)_{0,0} + (\Gamma^1)_{1,1}}. \label{eq:GC-def}
\end{equation}
GC analyses were performed at an order $k = 2$. Analyses at $k = 1$ yielded however fully analogous performance (not shown).

We note that the same pre-processing used for TE is also adopted for all the other analyses. The same holds for conditioning on the value of the average fluorescence~$g_n$, which can be applied simply by only including the subset of samples in which $g_{t}<\tilde{g}$.

\subsection*{Hubs of (causal) connectivity}

Connectivity in reconstructed networks is often inhomogeneous, and groups of nodes with tighter internal connectivity are sometimes visually apparent (see e.g. reconstructed topologies in Figure~\ref{fig:state_dependency}C). We do not attempt a systematic reconstruction of network communities~\cite{Fortunato:2010iw}, but we limit ourselves to the detection of ``causal sink'' nodes~\cite{Seth:2005wp}, which have a larger than average in-degree. We define this property in terms of the sum of TE from all other nodes to one particular node ($\sum_j \text{TE}_{j \rightarrow i}$), choosing the top~20 nodes for each particular network as selected ``hub nodes''.

We then analyze the dynamics of these selected hub nodes and of their neighbors. Specifically we define as $C$ the subgraph spanned by a given hub node and by its first neighbors. We analyze then the cross-correlogram of the average fluorescence of a given group~$C$ with the average fluorescence of the whole culture:
\begin{equation}
    \psi(\tau) = \mbox{corr} \left( \Delta \langle F^{\tt{sc}}(t+\tau) \rangle_{i \epsilon C}, \, \Delta \langle F^{\tt{sc}}(t) \rangle_i
    \right).
\end{equation}
The $\Delta$-notation indicates that we correlate discretely differentiated average
fluorescence time series, rather than the average time series themselves. Indeed,
cross-correlograms for these differentiated time series are well modeled by a Gaussian
functional form, due to the slow change of the averaged fluorescence compared to the
sampling rate.
% (see Fig.~S3).

Therefore, we fit a Gaussian to the cross-correlogram $ \psi(\tau)$:
\begin{equation}
    \psi_{fit}(\tau) = A_C \exp \left\{ - \left( \frac{\tau_C - \tau}{\sigma_C} \right)^2
    \right\},
\end{equation}
determining thus a cross-correlation amplitude~$A_C$, a cross-correlation peak lag~$\tau_C$ and the standard deviation~$\sigma_C$.

The cross-correlation peak lag~$\tau_C$ indicates therefore whether nodes in a given local hub neighborhood $C$ fire on average earlier or later than other neurons in the network.

Relative strength of synchrony within a local hub neighborhood $C$ can be analogously
evaluated by computing XCs, as defined in Eq.~\eqref{eq:Xcorr-def}, for all the links
within $C$ and comparing it with peak XCs over the entire network.

\subsection*{Experimental preparation}

Primary cultures of cortical neurons were prepared following standard procedures~\cite{Papa:1995uu,Soriano:2008p747}. Cortices were dissected from Sprague-Dawley embryonic rat brains at~19 days of development, and neurons dissociated by mechanical trituration. Neurons were plated onto 13 mm glass cover slips (Marienfeld, Germany) previously coated overnight with 0.01\% Poly-l-lysine (Sigma) to facilitate cell adhesion. Neuronal cultures were incubated at 37~$^{\circ}$C, 95\% humidity, and 5\%~$\text{CO}_2$ for 5~days in plating medium, consisting of 90\% Eagle's MEM ---supplemented with 0.6\%~glucose, 1\% 100X~glutamax (Gibco), and 20~$\mu$g/ml gentamicin (Sigma)--- with 5\%~heat-inactivated horse serum (Invitrogen), 5\%~heat-inactivated fetal calf serum (Invitrogen), and 1~$\mu$g/ml B27 (Invitrogen). The medium was next switched to changing medium, consisting of of 90\%~supplemented MEM, 9.5\%~heat-inactivated horse serum, and 0.5\%~FUDR (5-fluoro-deoxy-uridine) for 3~days to limit glia growth, and thereafter to final medium, consisting of 90\%~supplemented MEM and 10\%~heat-inactivated horse serum. The final medium was refreshed every 3~days by replacing the entire culture well volume. Typical neuronal densities (measured at the end of the experiments) ranged between 500 and 700~neurons/mm$^2$.  
\JordiRev{Cultures prepared in these conditions develop connections within 24 hours and show spontaneous activity by day \emph{in vitro} (DIV)~3-4 \cite{Wagenaar:2006fo,Soriano:2008p747,Cohen:2008cv}. GABA switch, the change of GABAergic response from excitatory to inhibitory, occurs at DIV~6-7  \cite{Soriano:2008p747,Cohen:2008cv}.}

Neuronal activity was studied at day \emph{in vitro} (DIV)~9-12. Prior to imaging,
cultures were incubated for 60~min in pH-stable recording medium in the presence of 0.4\%
of the cell-permeant calcium sensitive dye Fluo-4-AM (Invitrogen). 
\JordiRev{
Recording solution includes (in mM): $128\mathrm{~NaCl}$, $4\mathrm{~KCl}$, $1\mathrm{~CaCl}_2$, $1 \mathrm{~MgCl}_2$, 10 glucose, and 10 HEPES; pH titrated to 7.4 and osmolarity to $320 \mathrm{~mOsm}$ with $45 \mathrm{~mM}$ sucrose.
}
The culture was washed
off Fluo-4 after incubation and finally placed in a chamber filled with fresh recording
medium. The chamber was mounted on a Zeiss inverted microscope equipped with a
5X~objective and a 0.4X~optical zoom.

Neuronal activity was monitored through high-speed fluorescence imaging using a Hamamatsu
Orca Flash 2.8 CMOS camera attached to the microscope. Images were acquired at a speed of
100~frames/s (i.e. 10~ms between two consecutive frames), which were later converted to a
20~ms resolution using a sliding window average to match the typical temporal resolution of such recordings~\cite{Feinerman:2005ka,Sasaki:2007p2717,Feinerman:2008vc,Vogelstein:2009p3029,Mishchencko:2009p4301}.
The recorded images had a size of 648~$\times$~312 pixels with 256 grey-scale levels, and a
final spatial resolution of 3.4~$\mu$m/pixel. This settings provided a final field of
view of 2.2~$\times$~1.1~mm$^2$ that contained on the order of 2000~neurons.
\JordiRev{
Activity was finally recorded as a long image sequence of 60~minutes
in duration. We verified that the fluorescence signal remained stable during the recording, as shown 
in Supplementary Figure~S8A, and we did not observe neither photo-bleaching of the calcium probe nor photo-damage
of the neurons.
}

Before the beginning of the experiment, inhibitory synapses were fully blocked with
\JordiRev{$40~\mu\mathrm{M}$ bicuculline}, a GABA$_A$ antagonist, so that activity was solely driven by
excitatory neurons. 
\JordiRev{
Since cultures were studied after GABA switch, the blockade of inhibition resulted in an increase of the
fluorescence amplitude, which facilitated the detection of neuronal firing, as illustrated in Fig.~S8B.
}

The image sequence was analyzed at the end of the experiment to identify all active
neurons, which were marked as regions of interest~(ROIs) on the images. The average
grey-level on each ROI along the complete sequence finally provided, for each neuron, the
fluorescence intensity as a function of time. Each sequence typically contained on the
order of a hundred bursts. 
\JordiRev{
Examples of recorded fluorescence signal for individual neurons are
shown in Fig.~\ref{fig:raw_data_vs_simulations}B.
}

\subsection*{Analysis of experimental recordings}

The fluorescence data obtained from recordings of neuronal cultures was analyzed
following exactly the same procedures used for simulated data (e.g. processed in a
pipeline including discrete differentiation, TE or other metrics evaluation, ranking, and
final thresholding to maintain the top 10\% of connections).

Due to the lack of knowledge of the ground-truth topology, optimal conditioning level
cannot be known. However, based on the similarity between experimental and simulated
distributions of calcium fluorescence, we select a conditioning level such to exclude the
high fluorescence transients associated to fully-developed bursting transients while
keeping as many data points as possible. Concretely, this is achieved by taking a
conditioning level equal to approximately two standard deviations above the mean of a
Gaussian fit to the left peak of the fluorescence histogram. Such a level coincides with
the point where, when gradually increasing the conditioning level, the reconstructed
clustering index reaches a plateau, i.e. matches indicatively the upper limit of range II
in Figure~\ref{fig:state_dependency}.

To check for robustness of our reconstruction, we generated alternative reconstructions
based on different conditioning levels. For the selected conditioning value, and for both
the experimental datasets analyzed (Figs.~\ref{fig:real_data}~and~S7), we verified that
the inferred topological features, including notably the average clustering coefficient
and connection distance, were stable in a range centered on the selected conditioning
value and wide as much as approximately two standard deviations of the fluorescence
distribution.

To identify statistically significant non-random features of the real cultured networks
in exam, we compared the reconstructed topology to two randomizations.

A first one consisted in a complete randomization that preserved only the total number of
connections in the network, but scrambled completely the source and target nodes. The
resulting random ensemble of graphs was an Erd\"os-R\'enyi ensemble (see, e.g.
\cite{Newman:2001un}) in which each possible link exists with a uniform probability of
connection $p = C/(N(N-1))$, where~$C$ is the total number of connections in the
reference reconstructed network.

A second partial randomization preserved the in-degree distributions only, and was
implemented by shuffling the entries of each row of the reconstructed adjacency matrix,
internally row-by-row. In this way, the out-degrees of each node were preserved. In both
randomization processes, we disallowed diagonal entries.

For both randomizations we calculated the in-degree, the distance of connections and the
full clustering index for each node, leading to distributions of network topology
features that could be compared between the reconstructed network and the randomized
ensembles, to identify significant deviations from random expectancy.

% Do NOT remove this, even if you are not including acknowledgments
\section*{Acknowledgments}

The authors would like to thank Elisha Moses, Cyrille Zbinden, Liam Paninski and Elad Schneidman for stimulating discussions. This research was supported by the German Ministry for Education and Science~(BMBF) via the Bernstein Center for Computational Neuroscience~(BCCN) G\"ottingen (Grant No.~01GQ0430) and the Minerva Foundation, M\"unchen, Germany. JS acknowledges the financial support from the Ministerio de Ciencia e Innovaci\'on (Spain) under projects FIS2009-07523 and FIS2010-21924-C02-02, and the Generalitat de Catalunya under project 2009-SGR-00014.

%\section*{References}
% The bibtex filename
\bibliography{references}

\section*{Figure Legends}

\vspace{1em}\noindent
\textbf{Figure 1. Network activity in simulation and experiments.}
\textbf{A}~Bright field image \OlavRev{(left panel)} of a region of a neuronal culture at day \emph{in vitro} 12,  togther with its corresponding fluorescence image \OlavRev{(right panel)}, integrated over~200 frames. Round objects are cell bodies of neurons.
\textbf{B}~Examples of real~(left) and simulated~(right) calcium fluorescence time series, vertically shifted for clarity. \textbf{C}~Corresponding averages over the whole population of neurons. \textbf{D}~Distribution of population averaged fluorescence amplitude for the complete time series, for a real network~(left) and a simulated one~(right).

\vspace{2em}\noindent
\textbf{Figure 2. Independence of network dynamics from clustering coefficient.} \textbf{A}~Examples of spike raster plots for three networks with different clustering coefficients (non-local clustering ensemble), showing that their underlying dynamics are similar. \textbf{B}~Histograms of the inter-burst intervals~(IBIs), with the vertical lines indicating the mean of each distribution. The insets illustrate the amount of clustering by showing the connectivity of simple networks that have the same clustering coefficients as the simulated ones.

\vspace{2em}\noindent
\textbf{Figure 3. Dependence of the \DemianRev{directed functional} connectivity on the dynamical state.} \textbf{A}~The distribution of averaged fluorescence amplitudes is divided into seven fluorescence amplitude ranges. The \DemianRev{functional} connectivity associated to different dynamical regimes is then assessed by focusing the analysis on specific amplitude ranges. \textbf{B}~Quality of reconstruction as a function of the average fluorescence amplitude of each range. The blue line corresponds to an analysis carried out using the entire data sampled within each interval, while the red line corresponds to an identical number of data points per interval. \textbf{C}~Visual representation of the reconstructed network topology (top 10\% of the links only), together with the corresponding ROC curves, for the seven dynamical regimes studied. \OlavRev{Edges marked in green are present in both the reconstructed and the real topology, while edges marked in red do not match any actual structural link. Reconstructions are based on an equal number of data points in each interval, therefore reflecting the equal sample size performance (red curve) in panel B.} Interval~I corresponds to a noise-dominated regime; intervals~II to~IV correspond to inter-burst intervals with intermediate firing rate and provide the best reconstruction; and intervals~V-VII correspond to network bursts with highly synchronized neuronal activity. Simulations were carried out on a network with local topology ($\lambda=0.25~\text{mm}$) and light scattering in the fluorescence dynamics. The results were averaged over~6 network realizations, with the error bars in~\textbf{B} and the shaded regions in~\textbf{C} indicating a 95\% confidence interval.

\vspace{2em}\noindent
\textbf{Figure 4. TE-based network reconstruction of non-locally clustered topologies.} \textbf{A}~ROC curve for a network reconstruction with generalized TE of Markov order~$k=2$, and with fluorescence data conditioned at $g < \tilde{g}=0.112$.
% The vertical line indicates the performance level at TP$_{10\%}$ (fraction of true positive at 10\% false positives).
The shaded area depicts the~95\% confidence interval, and is based on~6 networks. \textbf{B}~Comparison between structural (shown in blue) and reconstructed (red) network properties: clustering coefficients (top), degree distribution (center), and distance of connections (bottom). \textbf{C}~Reconstructed clustering coefficients as a function of the structural ones for different reconstruction methods. Non-linear causality measures, namely Mutual Information (MI, red) and generalized Transfer Entropy (TE, yellow), provide the best agreement, while a linear reconstruction method such as cross-correlation (XC, blue) fails, leading invariantly to an overestimated level of clustering. The error bars indicate~95\% confidence intervals based on 3~networks for each considered clustering level. All network realizations were constructed with a clustering index of~0.5, and simulated including light scattering artifacts in the fluorescence signal.

\vspace{2em}\noindent
\textbf{Figure 5. TE-based network reconstruction of locally-clustered topologies.} \textbf{A}~ROC curve for a network reconstruction with generalized TE~Markov order~$k=2$, with fluorescence data conditioned at $g < \tilde{g}=0.084$.
% The vertical line indicates the performance level at TP$_{10\%}$ (fraction of true positive at 10\% false positives).
The shaded area depicts the~95\% confidence interval based on~6 networks. \textbf{B}~Comparison between structural (top) and reconstructed (bottom) connectivity. For the reconstructed network (\OlavRev{after thresholding to retain the top 10\% of links
only}) true positives are indicated in green, and false positives in red. \textbf{C}~Comparison between structural (blue) and reconstructed (red) network properties: clustering coefficients (top), degree distribution (center) and distance of connections (bottom). \textbf{D}~Reconstructed length scales as a function of the structural ones for different reconstruction methods. The non-linear causality measures Mutual Information (MI, red) and generalized Transfer Entropy (TE, yellow) provide good reconstructions, while the linear cross-correlation (XC, blue) provides invariantly an underestimated length scale. The error bars indicate~95\% confidence intervals based on 3~networks per each considered length scale. All network realizations were constructed with a characteristic length scale~$\lambda=0.25~\text{mm}$, and simulations included light scattering artifacts.

\vspace{2em}\noindent
\textbf{Figure 6. Dependence of performance level on network clustering, conditioning level and light scattering artifacts.} The color panels show the overall reconstruction performance level, quantified by TP$_{10\%}$ (black, 0\%~true positives; white, 100\%~true positives), for different target ground-truth clustering coefficients and as a function of the used conditioning level. Five different reconstruction algorithms are compared: cross-correlation (XC), Granger Causality (GC) with order $k=2$, Mutual Information (MI), and Transfer Entropy (TE) with Markov orders $k=1, 2$. The top row corresponds to simulations without artifacts, and the bottom row to simulations including light scattering. Reconstructions with linear methods are performant only in the absence of light scattering artifacts. TE reconstruction with $k=2$ shows the best overall reconstruction performance, even with light scattering and for any target clustering coefficient. An optimal reconstruction is obtained in a narrow range surrounding the conditioning value of $\tilde{g}\simeq 0.2$.

\vspace{2em}\noindent
\textbf{Figure 7. Dependence of reconstruction quality on TE formulations and recording length.} \textbf{A}~ROC curves for network reconstructions of non-locally clustered (left panel) and locally-clustered topologies (right), based on three TE formulations: conventional~TE (blue), generalized TE with same bin interactions only (red) or including also optimal conditioning (yellow). The vertical lines indicate the performance level at TP$_{10\%}$, and provide a visual guide to compare the quality of reconstruction between different formulations. \textbf{B}~Decay of the reconstruction quality as measured by TP$_{10\%}$ for the two topology ensembles and for generalized TE with conditioning, as a function of the data sampling fraction~$s$. A full simulated recording of~1h in duration provides a data set of length $S_{\text{1h}}$, corresponding to a data sampling fraction of~$s=1$. Shorter recording lengths are obtained by shortening the full length time-series at a shorter length given by~$S'=S_{\text{1h}}/s$, with $s=1,...,19$. The insets show the same results but plotted as a function of~$S'$ in semi-logarithmic scale. For both \textbf{A} and \textbf{B}, the panels in the left column correspond to the non-locally clustered ensembles (cfr. Fig.~\ref{fig:one_reconstruction_illustrated-CC}), while the panels in the right column correspond to the locally-clustered ensemble (cfr. Fig.~\ref{fig:one_reconstruction_illustrated-Lambda}). Shaded regions and error bars indicate~95\% confidence intervals based on 6 networks.

\vspace{2em}\noindent
\textbf{Figure 8. Network reconstruction of an {\textit in vitro} neuronal culture at DIV 12.} \textbf{A}~Example of TE reconstructed connectivity in a subset of 49~neurons (identified by black dots) in a culture with $N=1720$ marked neurons (regions of interest) in the field of view, studied at day \emph{in vitro}~12. \JordiRev{Only the top~5\% of connections are retained in order to achieve, in the final reconstructed network, an average connection degree of $100$ (see \textit{Results})}. \textbf{B}~Properties of the network inferred from TE reconstruction method (top panels) compared to a cross-correlation (XC) analysis (bottom panels). The figure shows reconstructed distributions for the in-degree (left column), the connection distance (middle column), and the clustering coefficient (right column). In addition to the actual reconstructed histograms (yellow), distributions for randomized networks are also shown. Blue color refers to complete randomizations that preserves the total number of connections only, and red color to partial randomizations that shuffle only the target connections of each neuron in the reconstructed network.

\section*{Supporting Figure Legends}

\vspace{1em}\noindent
\textbf{Supplementary Figure 1: \JordiRev{Dependence of reconstructed degree and clustering coefficient on} \DemianRev{the fraction of included links. We report here analyses for three representative} \JordiRev{non-locally clustered networks with clustering coefficients 0.1, 0.3 and 0.5 (as in Fig.~\ref{fig:CCnets_all_look_alike}).}} \OlavRev{\textbf{A} The average degree, by construction, varies linearly with the fraction of included links (or, shortly, threshold). \textbf{B} The reconstructed clustering coefficient show a steep rise at low threshold values, and describe broad hill-like profiles between approximately 5\% and 10\% of reconstructed links, which peaks around CC values matching approximately the ground truth clustering value of the network. For higher values of the threshold, the dependency of the reconstructed clustering coefficient against threshold becomes closer to the linear. Note that by definition of the ``full'' clustering coefficient, here measured (see \emph{Methods} section), a random network would show a perfect linear correlation.}

\vspace{2em}\noindent
\textbf{Supplementary Figure 2: \DemianRev{Hubs of functional connectivity}}.
\OlavRev{\textbf{A}~Plot of the spatial position of neurons of an example simulated network, highlighting functional connectivity hubs (red dots) for each of the dynamic intervals I-VII depicted in Fig.~\ref{fig:state_dependency}.}
\textbf{B}~For the corresponding dynamical regimes, the average cross-correlation between calcium signals of first-neighbor nodes of \DemianRev{functional} connectivity hubs (blue bars, see \textit{Materials and Methods}) is compared to the average cross-correlation between nodes inside the latter local groups and the rest of the population (green). Significance analysis is carried out using a Mann-Whitney test (n.s.: not significant). The results show that the degree of synchronization within the neighborhood of \DemianRev{functional} connectivity hubs is significantly higher than across groups.
\textbf{C}~Crosscorrelogram time-shift of the average activity of the same local groups with respect to the whole network average (see \emph{Methods}, cfr. also Fig.~S3). Negative shifts denote ``bursting earlier'' and positive shifts ``bursting later''. The analysis shows that the calcium fluorescence signal in the neighborhood of \DemianRev{functional} connectivity hubs for the dynamic ranges II-IV is ``leading'', i.e. has a negative lag in relation to the average population. Such \DemianRev{functional} connectivity hubs correspond therefore to foci of burst initiation.

\vspace{2em}\noindent
\textbf{Supplementary Figure 3: \OlavRev{State dependency in simulated and real data.}} \JordiRev{Two experiments are considered, ``A'' being the one at DIV 12 (Fig.~\ref{fig:real_data}) and~``B'' corresponding to the one at DIV 9 (Fig.~S7), together with a simulated data set, corresponding to the network from the local clustering ensemble (studied in Fig.~\ref{fig:one_reconstruction_illustrated-Lambda}). For the three cases we separate four ranges of fluorescence signal and identify accordingly four dynamical regimes (encoded by different colors, top row). We compute then, in each one of these regimes, the clustering index (central row) and the average connection distance (bottom row).
While the exact reconstructed values of the clustering index and the average connection distance are of course different, all three datasets share the main qualitative features, like the fact that the reconstructed clustering index is peaked in range~``III'' (marked in yellow in the top panels) where their average connection distance is lowest.}

\vspace{2em}\noindent
\textbf{Supplementary Figure 4: Analysis of additional topological features.} \textbf{A}~In non-locally clustered topologies, the non-linear causality measures Mutual Information (MI, red) and Transfer Entropy (TE, yellow) correctly estimate the average connection distance, while a linear measure such as cross-correlation (XC, blue) fail to do so, invariantly under-estimating this distance, as for the locally clustered ensemble (cfr. Fig. \ref{fig:one_reconstruction_illustrated-Lambda}D). Note that non-locally clustered topologies are random by construction in terms of the spatial distribution of connections, and, therefore, they display a virtually identical average connection length, independent of the clustering coefficient. \textbf{B}~In locally-clustered topologies, the non-linear measures MI (red) and TE (yellow) also provide a good estimation of the resulting clustering coefficient, in contrast with cross-correlation, XC (blue), invariantly overestimating the clustering level, as for the non-locally clustered ensemble (cfr. Fig. \ref{fig:one_reconstruction_illustrated-CC}D). In both plots, the dashed line corresponds to perfect reconstruction.

\vspace{2em}\noindent
\textbf{Supplementary Figure 5: Dependence of performance on characteristic length scale, conditioning level and light scattering.} The color panels show the overall reconstruction performance level, quantified by TP$_{10\%}$ (black, 0\% true positives; white, 100\% true positives), for different target ground-truth clustering coefficients and as a function of the used conditioning level. Five different reconstruction algorithms are compared: cross-correlation (XC), Granger Causality (GC) with order $k=2$, Mutual Information (MI), and Transfer Entropy (TE) with Markov orders $k=1, 2$. The top row corresponds to simulations without artifacts, and the bottom row to simulations including light scattering. TE of order $k=2$ and with light scattering provides a fair reconstruction quality at any length scale for a conditioning value $\tilde{g}\simeq 0.1$. (Note that the scale bar of reconstruction performance is different from the one in Fig.~\ref{fig:2D-scans}.)

\vspace{2em}\noindent
\textbf{Supplementary Figure 6: \JordiRev{Positive precision curve (PPC) analysis.}} \JordiRev{
This figure provide an alternative description of the reconstruction performances as a function of the inclusion of different components of the algorithm, studied in Fig.~\ref{fig:reconstruction_dependencies}A. A definition of positive prevision curves is provided provided in \emph{Materials and Methods}, following Ref. \cite{Garofalo:2009jm}. The plots show the ``true-false ratio'' (TFR) as a function of the ``true-false sum'' (TFS), i.e. the number of reconstructed links, and describe indirectly the likelihood that an included link is actual true positive. Positive values correspond to a larger number of true than false positives among the reconstructed links. Similarly to Fig.~\ref{fig:reconstruction_dependencies}A, only generalized TE including both same-bin interactions and optimal conditioning robustly displays positive values of TFR in a broad range of TFS. Note that a threshold of 10\% top links (used for example at the bottom panel of Fig.~\ref{fig:one_reconstruction_illustrated-Lambda}B), provides a TFS value of 990. Shaded areas are 95\% confidence intervals across 6 networks.}

\vspace{2em}\noindent
\textbf{Supplementary Figure 7: Network reconstruction of an {\textit in vitro} neuronal culture at DIV 9.}
\OlavRev{\textbf{A}~Detail of example fluorescence time series of individual neurons (top panel) and population average of fluorescence (bottom panel). Both synchronous bursts and inter-burst modulations of the fluorescence baseline are easily detectable in these recordings.}
\textbf{B}~Example of TE reconstructed connectivity in a subset of 39~neurons (identified by black dots) in a culture with $N=1668$ marked neurons (regions of interest) in the field of view, studied at day \emph{in vitro}~9. Only the top~5\% of connections are shown.
\textbf{C}~Properties of the network inferred from TE reconstruction method (top panels) compared to a cross-correlation (XC) analysis (bottom panels). The figure shows reconstructed distributions for the in-degree (left column), the connection distance (middle column), and the clustering coefficient (right column). In addition to the actual reconstructed histograms (yellow), distributions for randomized networks are also shown. Blue color refers to complete randomizations that preserves the total number of connections only, and red color to partial randomizations that shuffle only the target connections of each neuron in the reconstructed network.
\vspace{2em}\noindent

\vspace{2em}\noindent
\textbf{Supplementary Figure 8: \OlavRev{Stability, development and axon length of neuronal cultures.}} \JordiRev{\textbf{A}~Typical calcium fluorescence recording of neuronal activity at DIV 12 along 1 hour, showing the stability of the fluorescence signal (blue) and the rich network activity.
% (B, left) Detail of fluorescence time series of individual neurons at DIV 9, illustrating the occurrence of individual firing events in combination with synchronized network bursts. (B, right) Corresponding raster plot for the first 15 min of recording and for 500 neurons (30\% of the neurons monitored).
% (C) Examples of individual neuronal activity at DIV 12 and corresponding raster plot (25\% of the neurons monitored along 15 min). For clarity, neuronal firings outside bursts are marked with arrowheads. All the data shown in panels (A)-(C) corresponds to experiments with inhibition blocked.
\textbf{B}~Comparison between the fluorescence signal before (bottom) and after (top) pharmacological blocking of GABAergic transmission at DIV 9. The increase in burst amplitude after blocking of GABA confirms that GABA switch has already occurred.
\textbf{C}~Direct visualization of GFP-transfected neurons at DIV 12. The highlighted axons are between 1 and 2~mm in length and reveal the existence of long range connections in early mature cultures.}

\vspace{2em}\noindent
\textbf{Supplementary Figure 9: Schematic representation of light scattering radius.} Light scattering artifacts are expected to affect fluorescence imaging data when neurons are sufficiently close to each other. In this example, neurons are indicated as black dots. For three particular neurons, the radius of the simulated light scattering artifact~$\lambda_{\text{sc}}=0.15~\text{mm}$ is shown as green circular area centered on each of these neurons.

\section*{Tables}
\begin{table}
    [!ht] \caption{\textbf{Synaptic weights used in the simulation.} Mean and standard deviation for the internal synaptic weights~$\alpha_{\text{int}}$, used in the simulation of 6 networks with a non-locally clustered ensemble (listed with ascending clustering coefficients~CC, and 6 networks with a locally-clustered ensemble (listed by ascending length scales~$\lambda$).}
    \begin{tabular}
        {|l|l|l|l|} \hline \multicolumn{2}{|c|}{Topological index} & $\langle \alpha_{\text{int}} \rangle$ (pA) & sd of $\alpha_{\text{int}}$ \\
        \hline CC & 0.1 & 6.604 & 0.146 \\
        & 0.2 & 6.156 & 0.124 \\
        & 0.3 & 5.719 & 0.054 \\
        & 0.4 & 5.361 & 0.113 \\
        & 0.5 & 5.274 & 0.067 \\
        & 0.6 & 5.214 & 0.209 \\
        \hline $\lambda$~(mm) & 0.25 & 5.207 & 0.171 \\
        & 0.5 & 6.241 & 0.166 \\
        & 0.75 & 6.481 & 0.150 \\
        & 1.0 & 6.556 & 0.230 \\
        & 1.25 & 6.505 & 0.158 \\
        & 1.5 & 6.519 & 0.113 \\
        \hline
    \end{tabular}
    \label{tab:synaptic_weights_used_for_simulation}
\end{table}

\newpage

\begin{flushleft}
{\Large
\textbf{Figures and Supporting figures}
}
\end{flushleft}

\vspace{6em}

% available sizes in PLoS research article (cm): 8.3 or 12.35 or 17.35

% --- SUPPLEMENTARY FIGURES ---
\begin{figure}
    [!ht]
    \begin{center}
        \includegraphics[width=12.35cm]{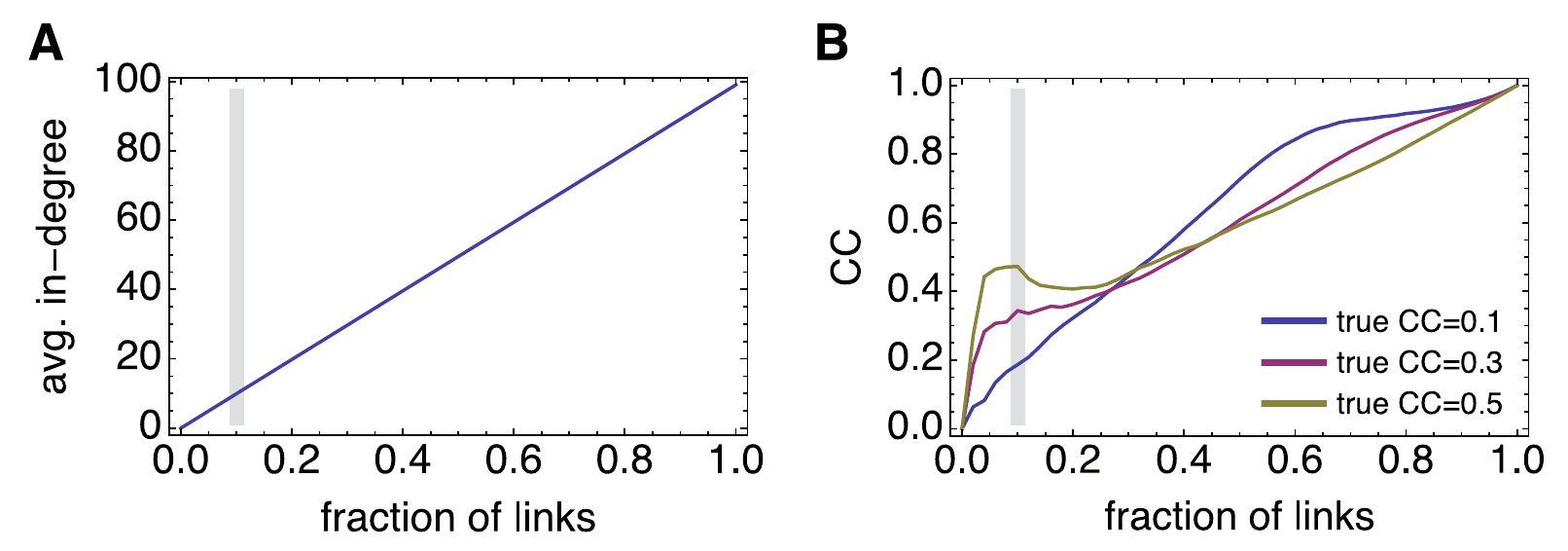}
    \end{center}
    \caption{ \textbf{[FIGURE S1]}} %\label{fig:threshold_effect}
\end{figure}

\begin{figure}
    [!ht]
    \begin{center}
        \includegraphics[width=17.35cm]{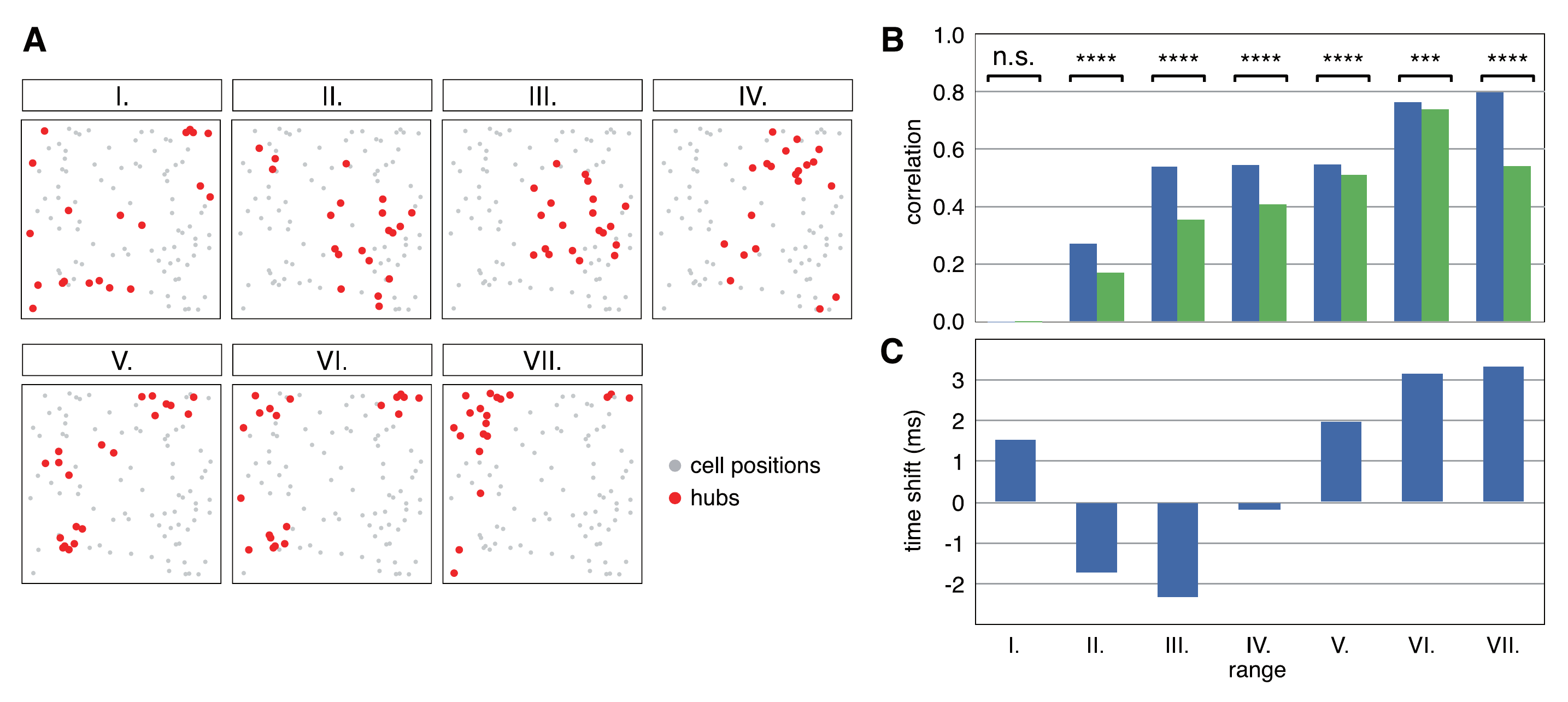}
    \end{center}
    \caption{ \textbf{[FIGURE S2]}} %\label{fig:community_barcharts}
\end{figure}

\begin{figure}
    [!ht]
    \begin{center}
        \includegraphics[width=17.35cm]{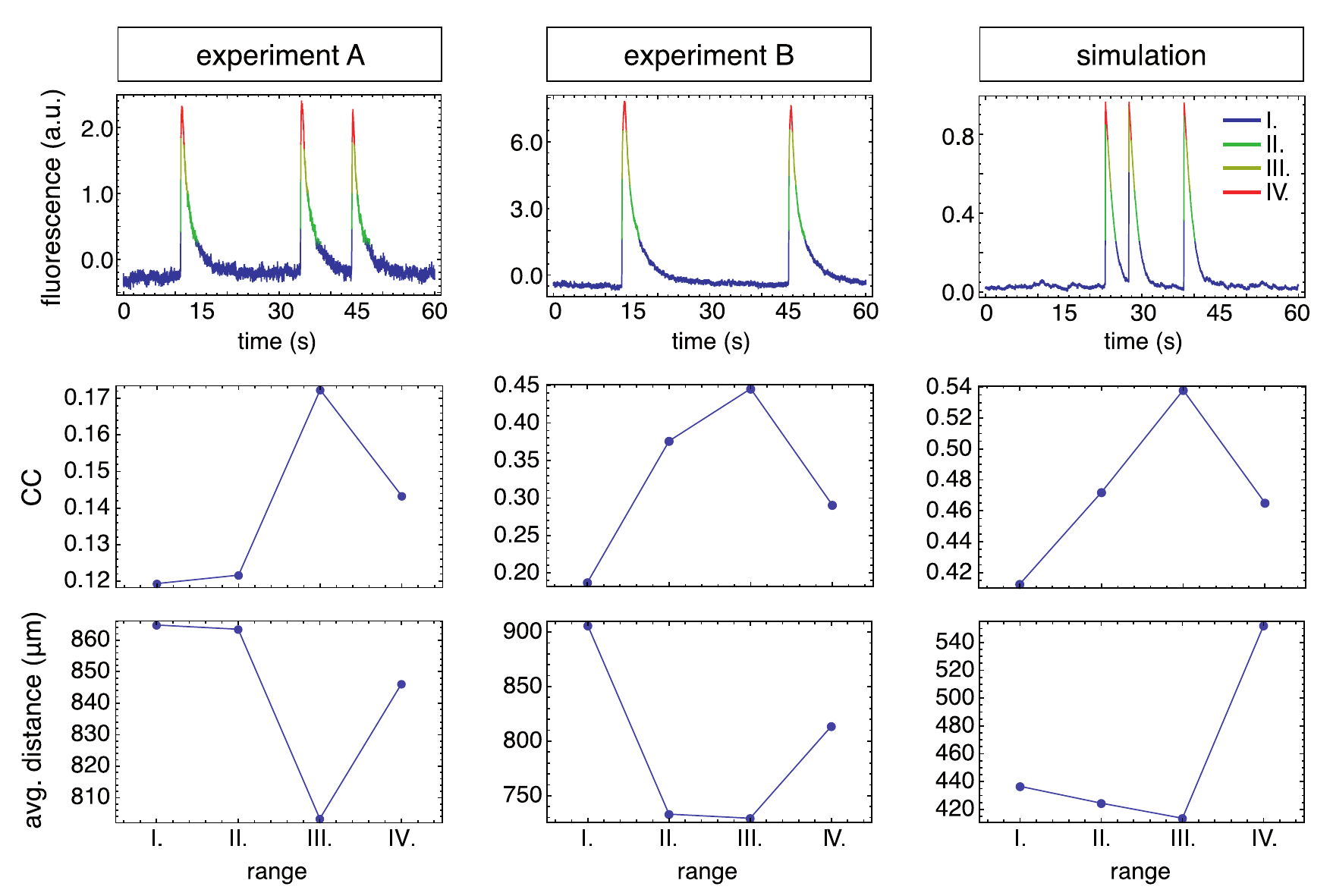}
    \end{center}
    \caption{ \textbf{[FIGURE S3]}} %\label{fig:state_dependency_real_data}
\end{figure}

\begin{figure}
    [!ht]
    \begin{center}
        \includegraphics[width=12.35cm]{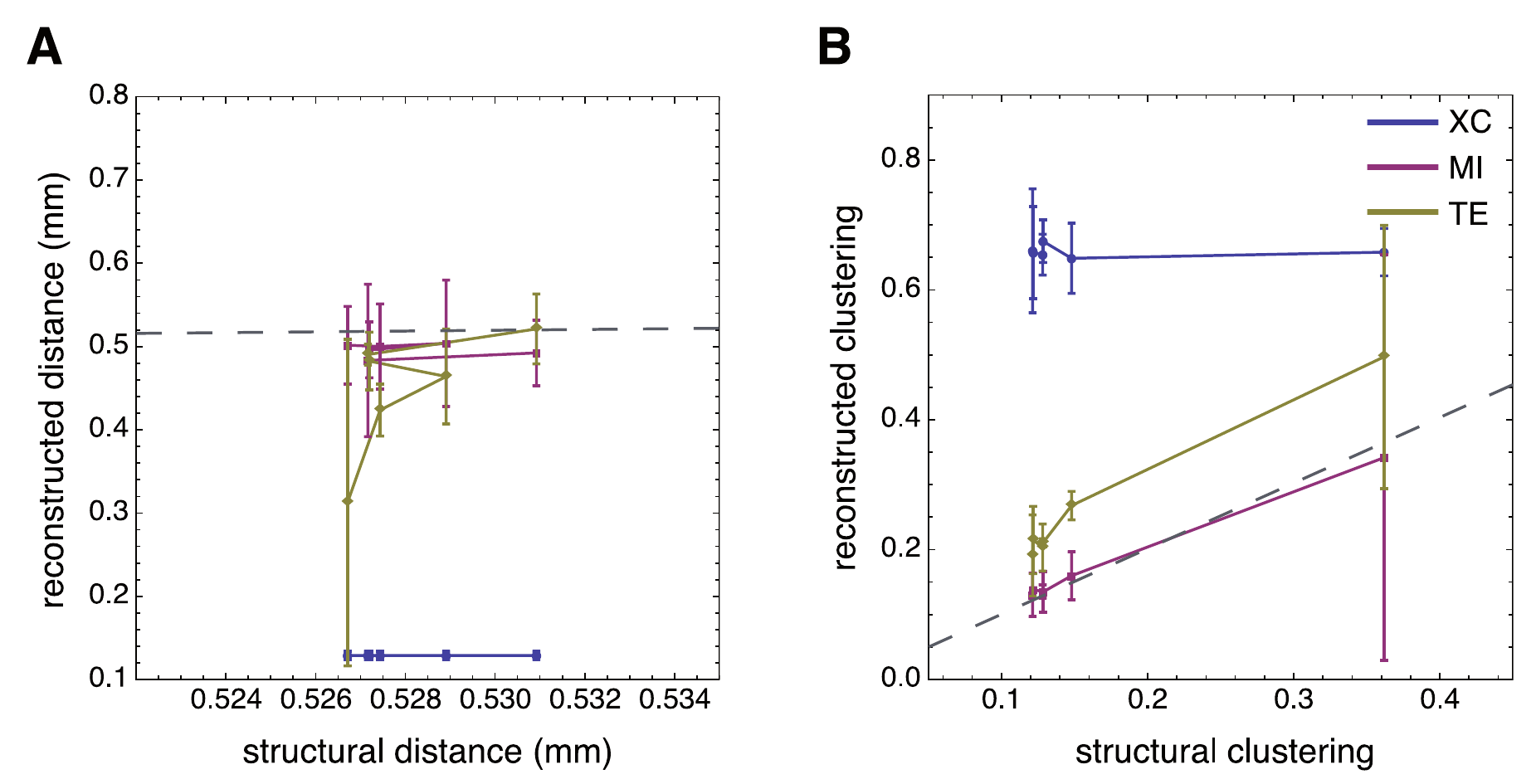}
    \end{center}
    \caption{ \textbf{[FIGURE S4]}} %\label{fig:real_vs_reconstructed_crossed}
\end{figure}

\begin{figure}
    [!ht]
    \begin{center}
        \includegraphics[width=17.35cm]{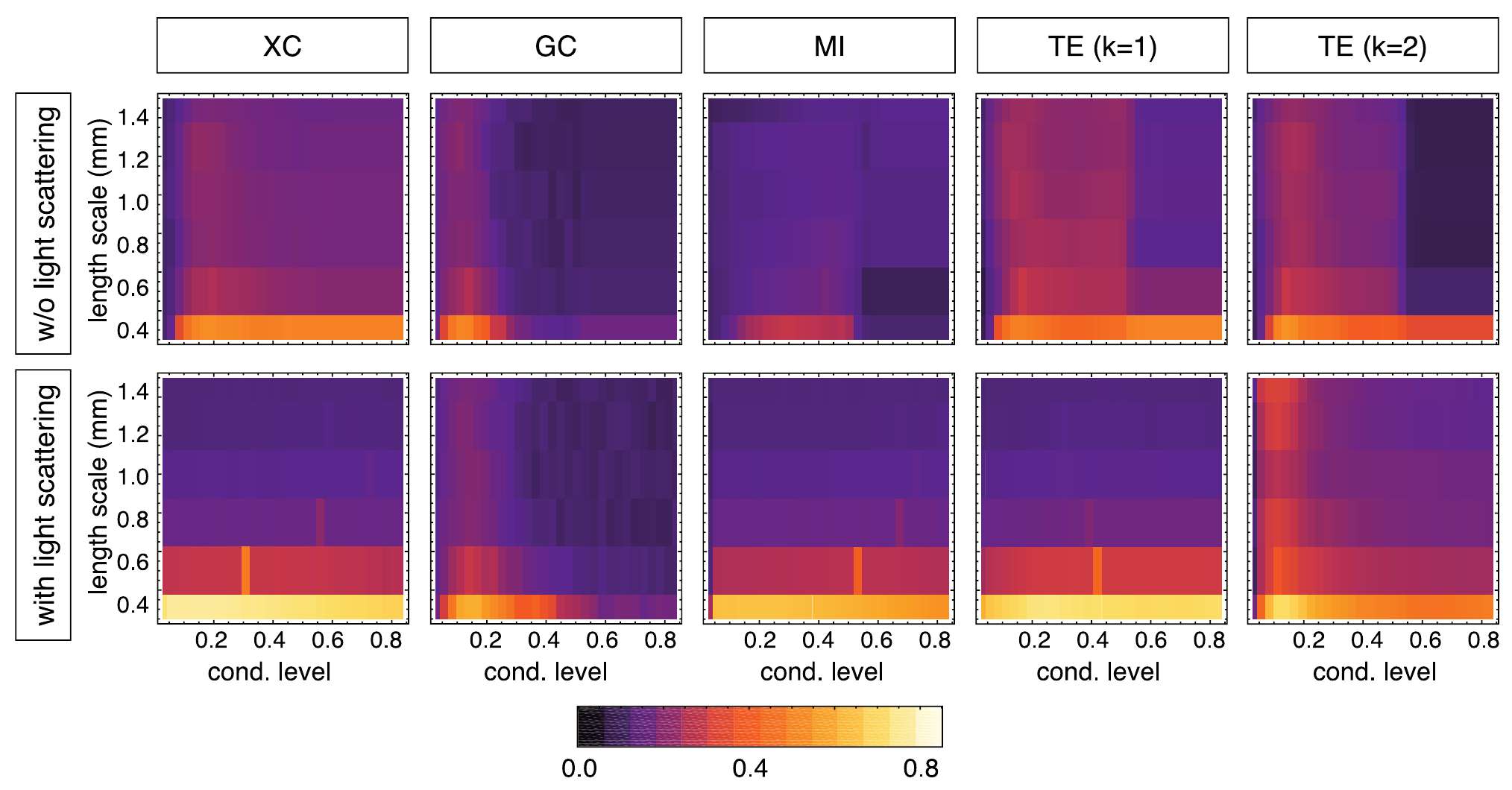}
    \end{center}
    \caption{ \textbf{[FIGURE S5]}} %\label{fig:2D-scans-Lambda}
\end{figure}

\begin{figure}
    [!ht]
    \begin{center}
        \includegraphics[width=12.35cm]{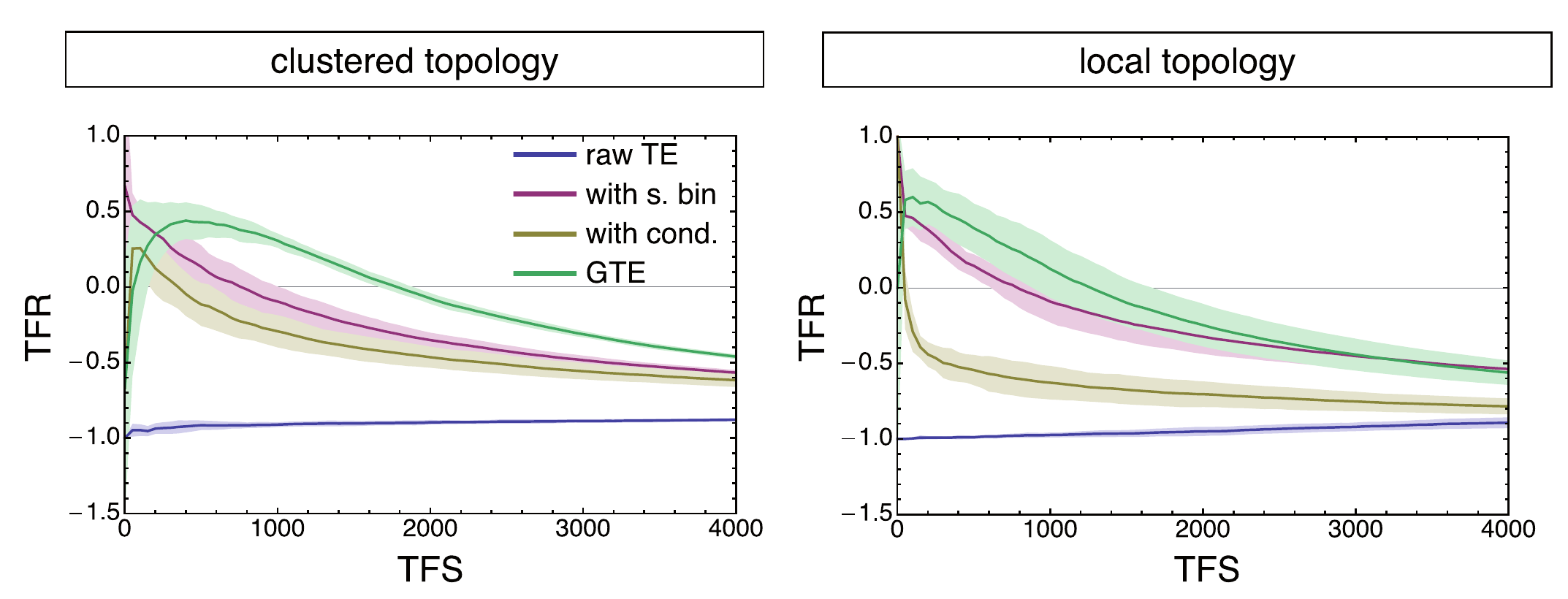}
    \end{center}
    \caption{ \textbf{[FIGURE S6]}} %\label{fig:ppcs}
\end{figure}

\begin{figure}
    [!ht]
    \begin{center}
        \includegraphics[width=17.35cm]{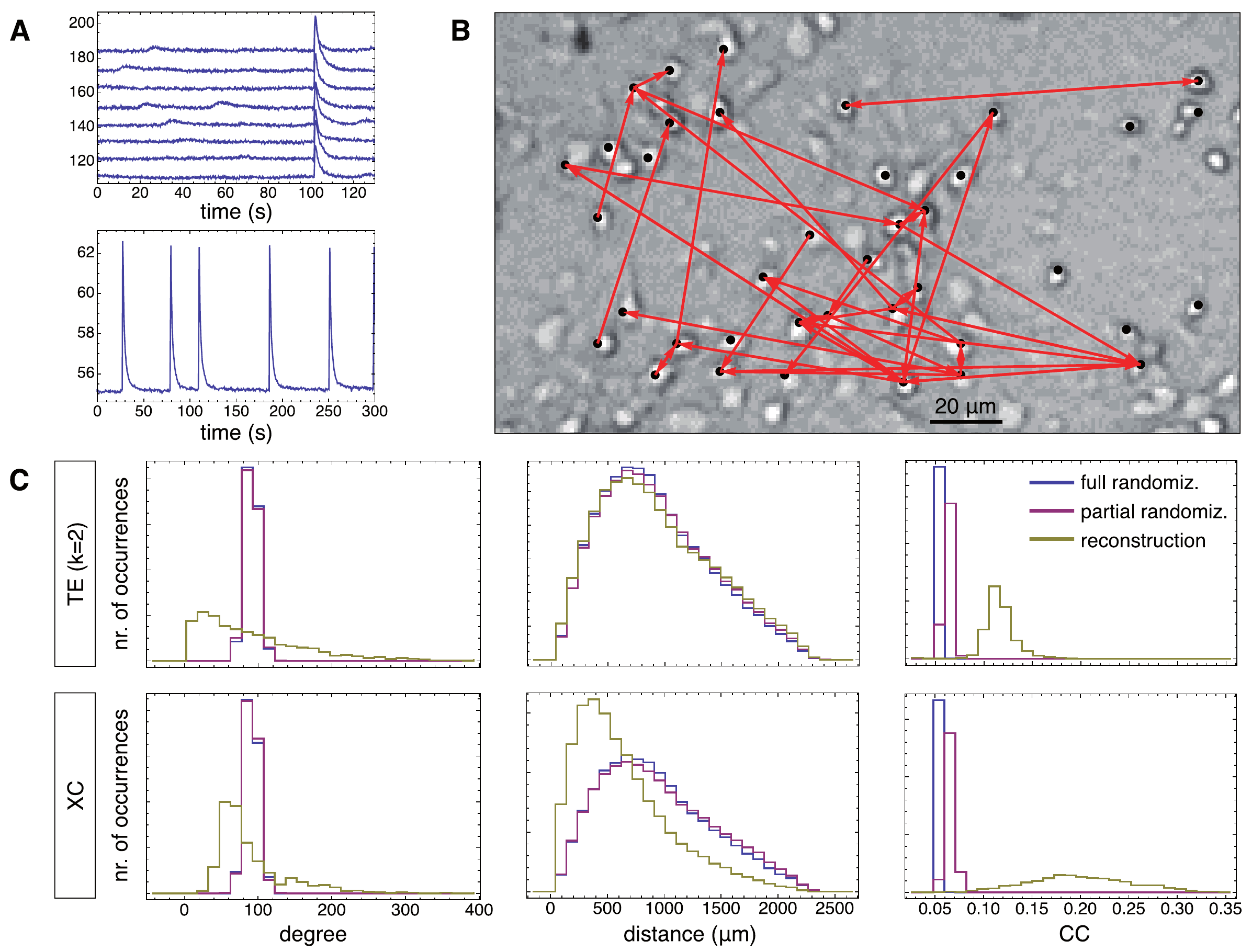}
    \end{center}
    \caption{ \textbf{[FIGURE S7]}} %\label{fig:more_real_data}
\end{figure}

\begin{figure}
    [!ht]
    \begin{center}
        \includegraphics[width=17.35cm]{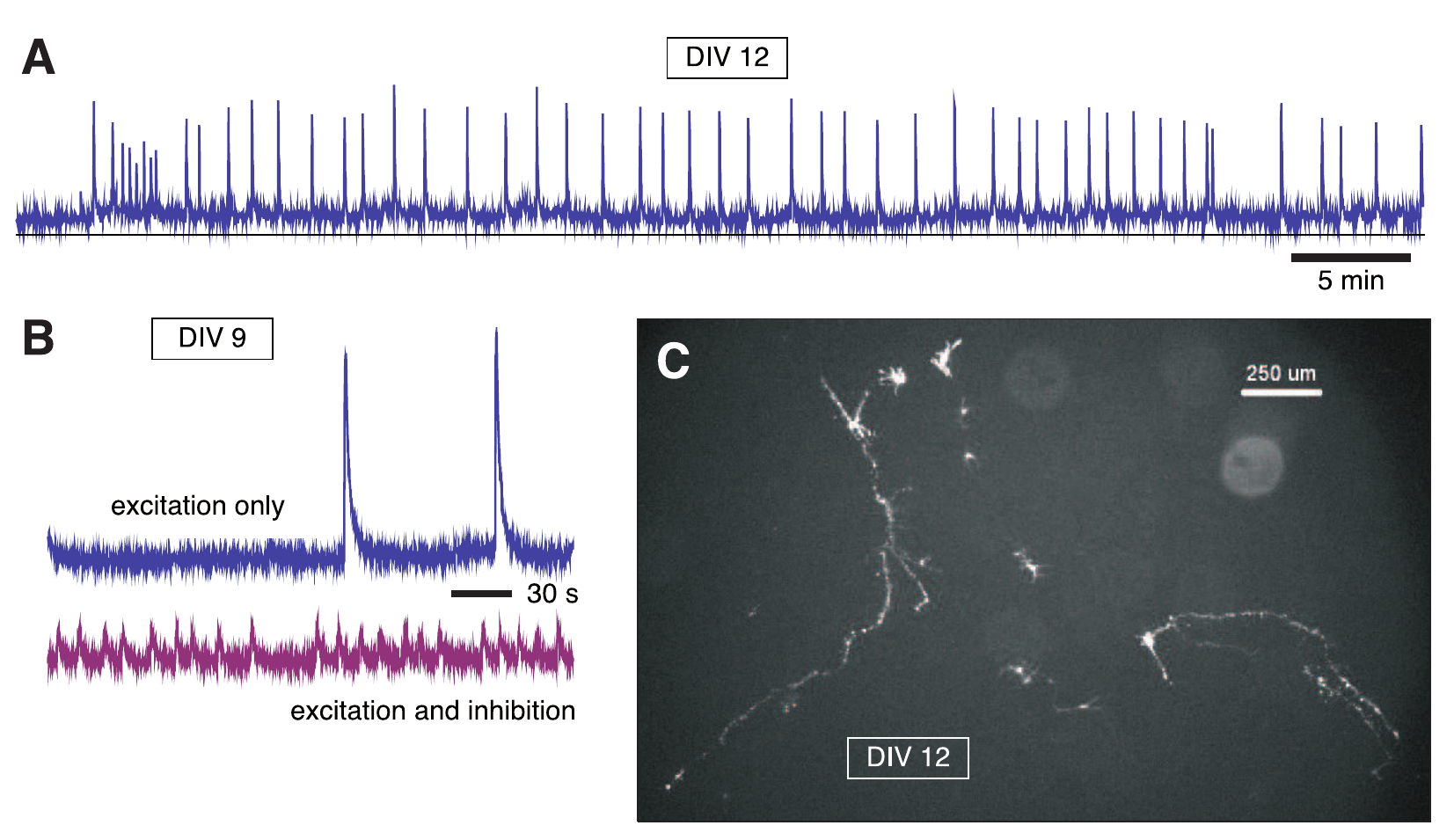}
    \end{center}
    \caption{ \textbf{[FIGURE S8]}} %\label{fig:Jordis_experimental_details}
\end{figure}

\begin{figure}
    [!ht]
    \begin{center}
        \includegraphics[width=8.3cm]{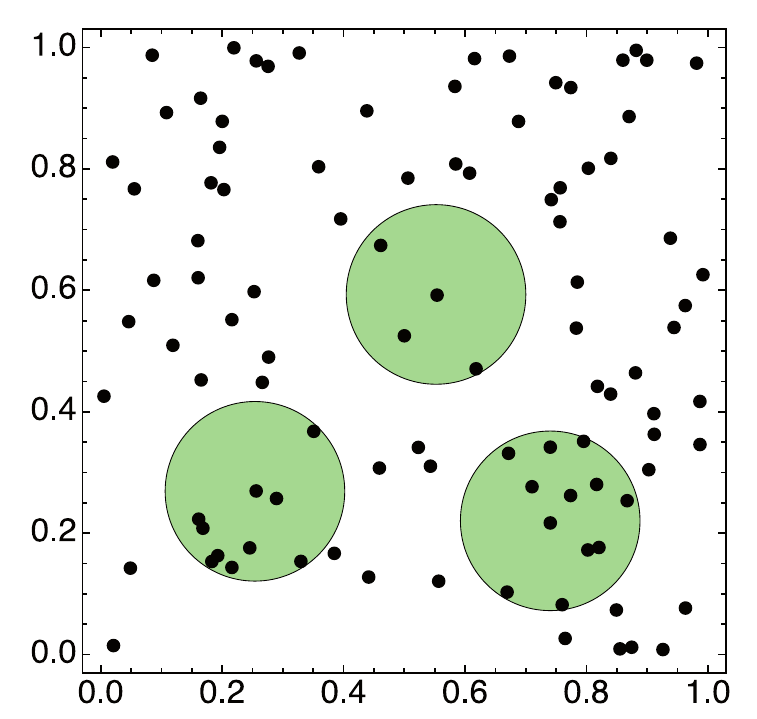}
    \end{center}
    \caption{ \textbf{[FIGURE S9]}} %\label{fig:scattering_radius}
\end{figure}

\end{document}